\documentclass[12pt,preprint]{aastex}

\usepackage{graphicx}
\usepackage{epstopdf}
\usepackage[usenames,dvips]{color}

\newcommand{\kms}{km s$^{-1}$}

\newcommand{\solar}{\ifmmode_{\sun}\;\else$_{\sun}\;$\fi}

\newcommand{\HI}{H$\,${\sc i}}

\newcommand{\nhi}{$N_{HI}$}

\newcommand{\rd}{$R_D$}
\newcommand{\rbr}{$R_{Br}$}

\newcommand{\coldens}{atoms cm$^{-2}$}

\begin{document}

\title{The Nature of Turbulence in the LITTLE THINGS Dwarf Irregular Galaxies}

\author{Erin Maier\altaffilmark{1,2},
Bruce G. Elmegreen\altaffilmark{3},
Deidre A. Hunter\altaffilmark{4},
Li-Hsin Chien\altaffilmark{1},
Gigja Hollyday\altaffilmark{4,5},
Caroline E. Simpson\altaffilmark{6}
}

\altaffiltext{1}{Department of Physics and Astronomy, Northern Arizona University, NAU Box 6010, Flagstaff, Arizona 86011 USA}
\altaffiltext{2}{Department of Physics and Astronomy, University of Iowa, 203 Van Allen Hall, Iowa City, IA 52242-1479 USA}
\altaffiltext{3}{IBM T. J. Watson Research Center, 1101 Kitchawan Road, Yorktown Heights, New York 10598 USA}
\altaffiltext{4}{Lowell Observatory, 1400 West Mars Hill Road, Flagstaff, Arizona 86001 USA}
\altaffiltext{5}{Department of Physics, University of Redlands, 1200 E.\ Colton Ave., Redlands, CA 92373-0999 USA}
\altaffiltext{6}{Department of Physics, Florida International University, CP 204, 11200 SW 8th St, Miami, Florida 33199 USA}

\begin{abstract}
We present probability density functions and higher order (skewness and kurtosis)
analyses of the galaxy-wide and spatially-resolved \HI\ column density distributions in
the LITTLE THINGS sample of dwarf irregular galaxies.
This analysis follows that of \citet{smc10} for the Small Magellanic Cloud.
About 60\% of our sample have galaxy-wide values of kurtosis that are
similar to that found for the Small Magellanic Cloud, with a range up to much higher values, and
kurtosis increases with integrated star formation rate. Kurtosis and skewness were
calculated for radial annuli and for a grid of 32 pixel $\times$ 32 pixel kernels
across each galaxy. For most galaxies, kurtosis correlates with skewness.
For about half of the galaxies, there is a trend of increasing
kurtosis with radius. The range of kurtosis and skewness values is modeled by
small variations in the Mach number close to the sonic limit and by
conversion of \HI\ to molecules at high column density.  The maximum \HI\ column
densities decrease with increasing radius in a way that suggests molecules are
forming in the weak field limit, where $H_2$ formation balances photodissociation
in optically thin gas at the edges of clouds.
\end{abstract}


\section{Introduction} \label{sec-intro}

Turbulence in the interstellar medium (ISM) has been proposed to be a key element in
determining the star formation laws of galaxies, 
especially in low gas density regimes such as those observed in the outer disks of spirals and in dwarf irregular (dIrr) galaxies.
Turbulent motions can both hinder star formation (SF) by preventing the collapse of star-forming clouds \citep[e.g.][]{maclow04} and help it by collecting gas into dense regions \citep{bruce93,bruce02}. There is also an expected relation between star formation activity and turbulence in the form of increased velocity dispersions due to the injection of energy from stars via stellar winds and, especially, supernovae explosions (SNe) \citep[e.g.][]{maclow04,tamburro09}. However, it is thought that SF cannot be the only driver of turbulence in the ISM, as observed velocity dispersions remain approximately constant across the disk, even in low density and low star-formation regions in both spirals \citep{tamburro09} and dwarfs \citep{hunter99,hunter01}. 

Previous studies of turbulence in the ISM in dwarf galaxies have focused largely on determining the source of the turbulence. 
\citet{stilp13}, for example, measured turbulence in the atomic hydrogen (\HI) of dwarf galaxies in the VLA-ANGST \citep{ott12} 
and THINGS \citep{walter08} surveys.  They measured the velocity dispersions from stacked and averaged individual 
line-of-sight \HI\ spectral profiles, either from the same radial annuli or from regions with the same SF surface density. 
They found that the measured velocity dispersions from regional ``superprofiles" correlate best with baryonic and \HI\ surface mass densities, 
indicating that some kind of gravitational instabilities are the most likely driving source. 
However, they found that none of the mechanisms they studied (which also included SF, 
magneto-rotational instability, and accretion-driven turbulence) could individually cause the observed level of turbulence in low SFR regions.

\citet{dutta09} used power spectra of the \HI\ in a sample of seven dwarf galaxies (and one spiral) to look at turbulence in the ISM. They found a correlation between higher SFR per unit area and steeper power-law slopes for four of their sample, but caution that confirmation with a larger sample of galaxies was needed. \citet{zhang12} used Fourier transform spatial power spectra of velocity slices of varying widths to examine the turbulence in the ISM of a sample of LITTLE THINGS dwarf galaxies \citep[Local Irregulars That Trace Luminosity
Extremes, The \HI\ Nearby Galaxy Survey,][]{lt12}\footnote[7]{Funded in part by the
National Science Foundation through grants AST-0707563, AST-0707426, AST-0707468, and
AST-0707835 to US-based LITTLE THINGS team members and with generous support from the
National Radio Astronomy Observatory.}. 
Observed power-law slopes in \HI-line intensity power spectra in the Milky Way \citep{cd83,green93,dickey01,khalil06} and Magellanic Clouds \citep{sl01,elmegreen01} are thought to be due to the presence of turbulence in the ISM \citep{zhang12}. 
The power-law behavior arises as driving sources input energy on large scales and the energy ``cascades" down with little dissipation to small scales, where it dissipates (this assumes a nonmagnetic incompressible medium) \citep{kolmogorov41}. The range of scales where little dissipation occurs is known as the ``inertial-range." 
Hydrodynamic galaxy simulations by \citet{bournaud10}, 
\citet{pilkington11},
\citet{combes12}, \citet{hopkins12},
and \citet{walker14}
show that large scale turbulence can be driven by gravitational instabilities, but that to maintain small-scale structures (clouds) in the ISM there needs to be a different source of energy --- this is thought to be from SNe. \citet{zhang12} find no correlation between the inertial-scale spectral indices and star-formation rates (SFR), so conclude that non-stellar sources are required to provide turbulence in the inertial range, or that turbulent structures are not related to driving sources. 

The relative importance of SNe and self-gravity for the generation of interstellar turbulence is not so clear in recent numerical simulations. While \cite{ibanez16} state that self-gravity is needed to explain the observed correlations between molecular cloud line-widths, sizes, and column densities \citep{heyer09}, \cite{padoan16} suggest that these correlations and others can result entirely from supernova pumping without self-gravity. \cite{kim14} and \cite{kim15} also find that star formation properties, scale heights, and velocity dispersions in galaxies can be explained by supernova feedback. Still, internal cloud feedback from pre-main sequence winds is probably not generating the observed cloud line-widths, even though there is enough energy and momentum in the winds to do this; self-gravity seems to be involved instead \citep{drabek16}.  \cite{liburkert17} and \cite{li17} also note that self-gravitational energy generation dominates turbulent energy decay in local molecular clouds. Most likely, a combination of supernova energy and gravitational energy drive most interstellar turbulence, with self-gravity dominating on galactic scales and in cloud interiors, and supernovae dominating on intermediate scales, i.e., smaller than the disk thickness and particularly in the lower density gas. Magnetic field instabilities should also contribute to turbulence in the lower density gas \citep[e.g.,][]{piontek07}.

By examining the change of slopes in power-spectra from different velocity widths, \citet{zhang12} also confirm the work of \citet{lp00} that shows that \HI\ power spectra made with different velocity width data trace different temperature phases of a multi-phase ISM. This means that observationally determining turbulent velocity spectral indices is complex and needs to be informed by models that include the multi-phase nature of the ISM. A multi-phase ISM consists of both a warm neutral medium (WNM) which is transonic, and a cold neutral medium (CNM) which is supersonic; so, unlike with the Kolmogorov nonmagnetic incompressible medium mentioned above, the ISM is compressible with both self-gravity and magnetic fields likely affecting turbulence. Higher sonic and Alfven Mach numbers indicate weaker magnetic fields and stronger shocks in the ISM. It has been suggested by \citet{krumholz05} that the star formation rate (SFR) is at least partly determined by the sonic Mach number in the ISM.

Numerical simulations by \citet{kraljic14} suggest that the change in slope
of the Schmidt-Kennicutt relation between gas surface density and 
SFR from the low gas density regime to high gas density regime is the result of a
transition from subsonic to supersonic turbulence \citep[see also][]{renaud12}. This
transition may also be coupled to the transition from inefficient to
efficient star formation. Dwarf irregular galaxies and the outer disks of spirals
that are largely in the low gas surface density regime are, in this model, expected to
have turbulence that is primarily subsonic as well as low efficiency star formation.
Furthermore, in these low gas density regimes turbulence may be a very important process
for bringing gas together into star-forming clouds since other processes believed to be
crucial in high density gas environments do not work at low gas densities
\citep[e.g.][]{bruce06}.

In this study we present an examination of the nature of the turbulence in the \HI\ gas
of a sample of nearby dIrr galaxies. (See \citealt{spirals} for a companion study of
spiral galaxies). In addition to constructing probability density functions (PDFs) of the
integrated \HI\ column densities, \nhi, we determine the third and fourth moments
(skewness and kurtosis) of PDFs for comparison to models \citep[][see also Konstandin et
al.\ 2012]{kowal07}. Skewness is a measure of the symmetry of the PDF distribution, and
kurtosis is a measure of the sharpness of the peak and extent of the tails of the
distribution. The analysis of higher order moments of \nhi\ distributions follows the
study of \citet{smc10} in which they determined the skewness and kurtosis values of the
\HI\ PDFs of the Small Magellanic Cloud (SMC) and inferred the Mach number of the gas from simulations.
\citet{smc10} found that the turbulence in the SMC is transonic, which implies that
the ISM is only weakly structured by turbulence compression.

The sample of dwarf irregulars used here is described briefly in Section
\ref{sec-data}. We present simple PDFs, the number of pixels as a function of
integrated \HI\ column density, as a function of radius in Section \ref{simplepdf}.
For the skewness and kurtosis analysis that follows, we discuss galaxy-wide values
and evidence for conversion of \HI\ to molecules in Section
\ref{sec-skwide}, variations with radius from measurements in annuli in Section
\ref{sec-skann}, and two-dimensional maps of values in Section \ref{sec-skkernel}.
In Section \ref{sec-discuss} we present models that reproduce the
skewness and kurtosis values that we find in terms of the Mach number of the gas
and an upper limit to the HI column density. A summary is in
Section \ref{sec-summary}.

\section{Sample and Data} \label{sec-data}

The sample of galaxies is taken from LITTLE THINGS \citep{lt12}. This is a multi-wavelength survey of
37 dIrr galaxies and 4 Blue Compact Dwarfs (BCDs; Haro 29, Haro 36, Mrk 178, VIIZw 403),
which builds on the THINGS project, whose emphasis was on nearby spirals
\citep{walter08}. The LITTLE THINGS sample contains dwarf galaxies that are relatively
nearby ($\leq$10.3 Mpc; 6\arcsec\ is $\leq$300 pc), that contain gas, the fuel for star
formation, and that cover a large range in dwarf galactic properties. The data used here
include \HI-line maps obtained with the Karl G.\ Jansky Very Large Array (VLA\footnote[8]{
The National Radio Astronomy Observatory is a facility of the National Science Foundation operated under cooperative agreement by Associated Universities, Inc.}). 
The \HI\ maps are characterized by high sensitivity ($\leq1.1$ Jy beam$^{-1}$ per
channel), high spectral resolution ($\leq$2.6 \kms), and high angular resolution
($\sim$6\arcsec). In addition we use 1516 \AA\ FUV images obtained with the {\it Galaxy
Evolution Explorer} satellite \citep[{\it GALEX};][]{martin05} as a tracer of star
formation over the past 100--200 Myrs. The galaxies and a few key parameters are listed
in Table \ref{tab-gal}.

\begin{figure}
\epsscale{0.95}
\plotone{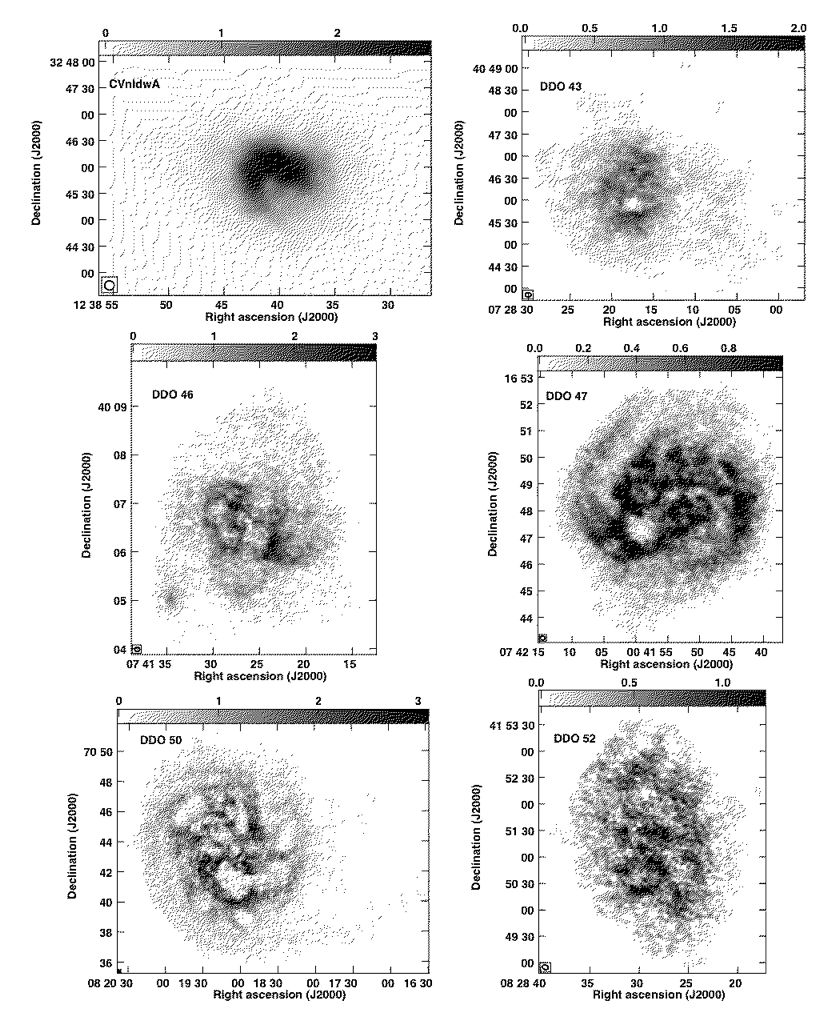}
\caption{
Robust-weighted integrated \HI\ (moment 0) column density maps of the LITTLE THINGS dwarf irregular galaxies.
See \citet{lt12} for details. The maps are in units of $10^{21}$ \coldens, except
for some galaxies where the units are $10^{18}$ \coldens\
(DDO 87, DDO 101, F564-V3, LGS3, M81dwA, SagDIG).
\label{fig-mom0r}}
\end{figure}

In this study we use the robust-weighted integrated \HI\ (moment 0) maps,
shown in Figure \ref{fig-mom0r}. This gives us
high angular resolution and is motivated by the 30-pc spatial resolution of the SMC used
by \citet{smc10} that we wish to compare to. However, only 4 galaxies (DDO 69, IC 10, IC
1613, WLM) have high enough angular resolution to allow a 30-pc spatial resolution. In
order to determine the effect of spatial resolution on the skewness and kurtosis values,
we computed these values at 4 different spatial resolutions: the original angular
resolution of the observations and maps smoothed, where possible, to 30 pc, 100 pc, and
200 pc beam-sizes. The original beam sizes are given in Table \ref{tab-results}.
A discussion of the effect of the varying resolutions on the results is presented below.

\section{Analysis} \label{sec-analysis}

The PDF of \HI\ column density gives information about the range of \nhi\ variations
and the limiting values. The range is related to the ISM phase structure, which includes cool clouds
and a warm intercloud medium, and it is related to substructure from self-gravity,
pressure-swept shells, and compressive turbulence. The limiting values extend from the
threshold of detection at low column density, which might coincide with lines of sight
through purely warm-phase gas, up to either cool \HI\ clouds or the
transition to molecules.  Here we study the \nhi\ PDF for each galaxy in four radial
bins, and for radii inside and outside the break radius of the exponential disks. We also
determine the 3rd and 4th moments of the PDFs for whole galaxies, as functions of radius,
and in square sub-region kernels. The results give some insight into the transition from
atomic hydrogen to molecules, and they indicate an approximate way to determine the Mach
number of the turbulence.

\subsection{Simple PDFs as a function of radius}
\label{simplepdf}

From the integrated \HI\ maps with the original beam-sizes given in Table
\ref{tab-results}, we computed the number of pixels as a function of column density,
\nhi. Pixels with \nhi\ $\le 15\times10^{18}$ cm$^{-2}$, which represents several
times the typical rms noise in the outskirts of the LITTLE THINGS galaxies, were eliminated
from each galaxy in order to ensure good S/N.
The pixel size is 1.5\arcsec\ for all galaxies except for DDO 216 and
SagDIG, which have 3.5\arcsec\ pixels. In order to remove the overall decline of \nhi\
with radius, which would otherwise dominate the PDF profile,
we normalized the individual values with the average determined radially
every $V$-band disk scale length, \rd.
Before making the average \nhi\ profiles, we clipped the integrated \HI\ maps
to remove pixels with values $\le 15\times10^{18}$ cm$^{-2}$, and only non-blank pixels were
included.
To find the radius of a given pixel in the plane of the
galaxy, we used the galactic center, inclination, and position angle of the major axis
determined from \HI\ kinematic fits \citep{oh15}, where available, or from the optical
$V$-band morphology \citep{he06}, otherwise.

\begin{figure}
\epsscale{0.95}
\plotone{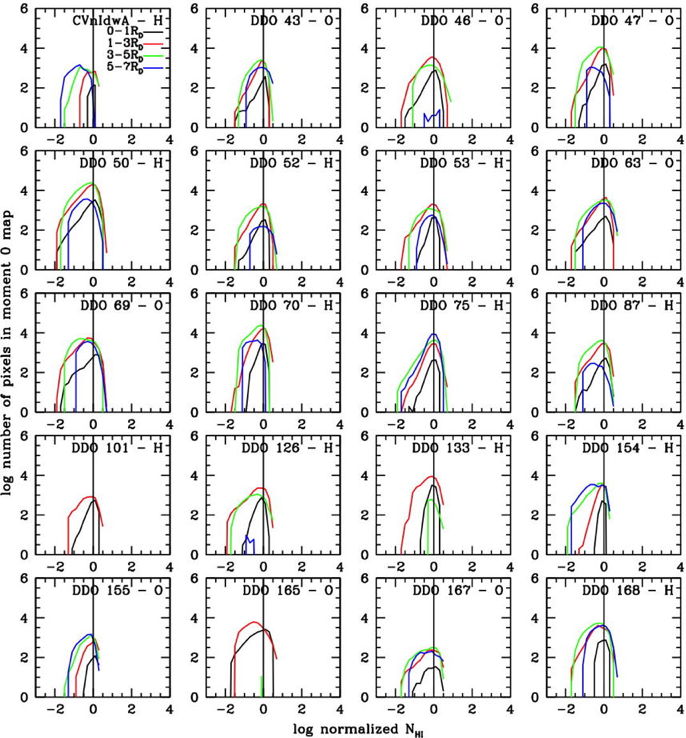}
\caption{\nhi\ PDFs as a function of distance from the center of the galaxy in units
of disk scale length \rd.
Wherever \HI\ kinematic fits to the velocity field are available \citep{oh15},
parameters from those fits -- galaxy center,
inclination angle, major axis position angle -- are used. Those galaxies are marked
with an ``H'' for ``\HI'' after the galaxy name. The rest of the galaxies used determinations
from $V$-band morphologies \citep{he06} and are denoted ``O'' for ``optical.''
Pixels with \nhi\ $\le 15\times10^{18}$ cm$^{-2}$ were eliminated to ensure good S/N.
\label{fig-pdfrd}}
\end{figure}

\begin{figure}
\epsscale{0.95}
\plotone{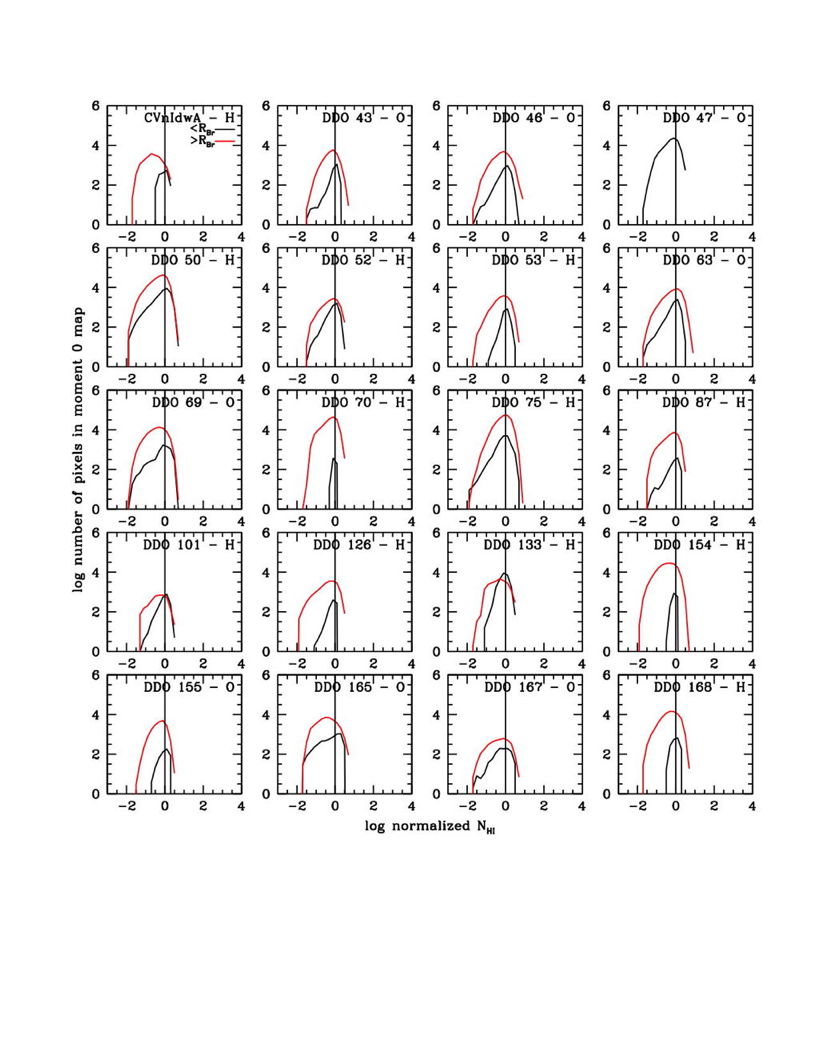}
\vskip -1.5truein
\caption{\nhi\ PDFs before and after breaks in the galaxy's $V$-band surface brightness profiles, \rbr,
as given by \citet{herrmann13}.
DDO 47 and DDO 210 do not exhibit breaks in their exponential surface brightness profiles,
so galaxy-wide PDFs are shown instead.
Wherever \HI\ kinematic fits to the velocity field are available \citep{oh15},
parameters from those fits -- galaxy center,
inclination angle, major axis position angle -- are used. Those galaxies are marked
with an ``H'' for ``\HI'' after the galaxy name. The rest of the galaxies used determinations
from $V$-band morphologies \citep{he06} and are denoted ``O'' for ``optical.''
Pixels with \nhi\ $\le 15\times10^{18}$ cm$^{-2}$ were eliminated to ensure good S/N.
\label{fig-pdfrbr}}
\end{figure}

Figure \ref{fig-pdfrd} shows the PDFs of normalized \nhi\  for each galaxy in
4 different radii bins: 0-1\rd, 1-3\rd, 3-5\rd, and 5-7\rd. The \rd\ were determined by
\citet{herrmann13} and are given in Table \ref{tab-gal}. Figure \ref{fig-pdfrbr} shows
the PDFs interior and exterior to the break, \rbr, in the $V$-band surface brightness
profile, if one is present. The break marks a change in slope in the profile, with most
bending downward and the exponential fall-off becoming steeper (Type II breaks), but a
few bending upward and becoming shallower with radius (Type III). The \rbr\ were measured
by \citet{herrmann13} and are also given in Table \ref{tab-gal}.

Some of the PDF shapes in Figure \ref{fig-pdfrd} are Gaussian in the central
regions (which means parabolic in these log-log plots) but they have sharper than
Gaussian drop-offs at low and high \nhi. The PDFs are similar in shape to those
made (not shown) without the cut-off of \nhi\ below $15\times10^{18}$ cm$^{-2}$, so
the lack of tails in the PDFs of Figure \ref{fig-pdfrd} is not due to this
cut-off. \citet{patra13} also saw a sharp drop off to higher \nhi\ in PDFs of the FIGGS
sample of dwarf galaxies. However, another common PDF shape in our sample is one in which
the fall-off to lower normalized \nhi\ is more gradual than to higher values, so the PDF
is not symmetrical on a log-log plot.

For radial trends in the PDFs, for many of the galaxies we see that the PDFs are similar at different
radii within a galaxy. This similarity for normalized \nhi\ implies that the absolute
\nhi\ distribution does not extend up to some fixed value such as a fixed molecular
transition threshold \citep[e.g.,][]{bigiel08} or fixed dust opacity. The upper limit to
the normalized column density, $N(HI)_{\rm max}$, is around a factor of 10 times the
average in most of these cases, which is much less than the expected upper limit to the
gas column density in general, especially since there is star formation.
That the upper limit to the density is lower than expected, however, is not due to a larger beam-size since some of these galaxies
have the highest linear resolution in our sample, as low as a 20 pc beam-size.
This upper limit seems instead
to imply that the disappearance of \HI\ emission at high density, which is presumably
from a transition to opaque \HI\ or molecules, occurs at a relative value of column
density, and is possibly driven by pressure and starlight. 
Although many of the galaxies have PDFs that are similar at different radii within the galaxy,
other galaxies in Figure
\ref{fig-pdfrd} have upper limits to the PDF functions that increase regularly with
increasing annulus.  This is partly because of the rejected pixels mentioned above; these
cases have a more constant value of the maximum column density in absolute terms.

\begin{figure}
\epsscale{0.95}
\plotone{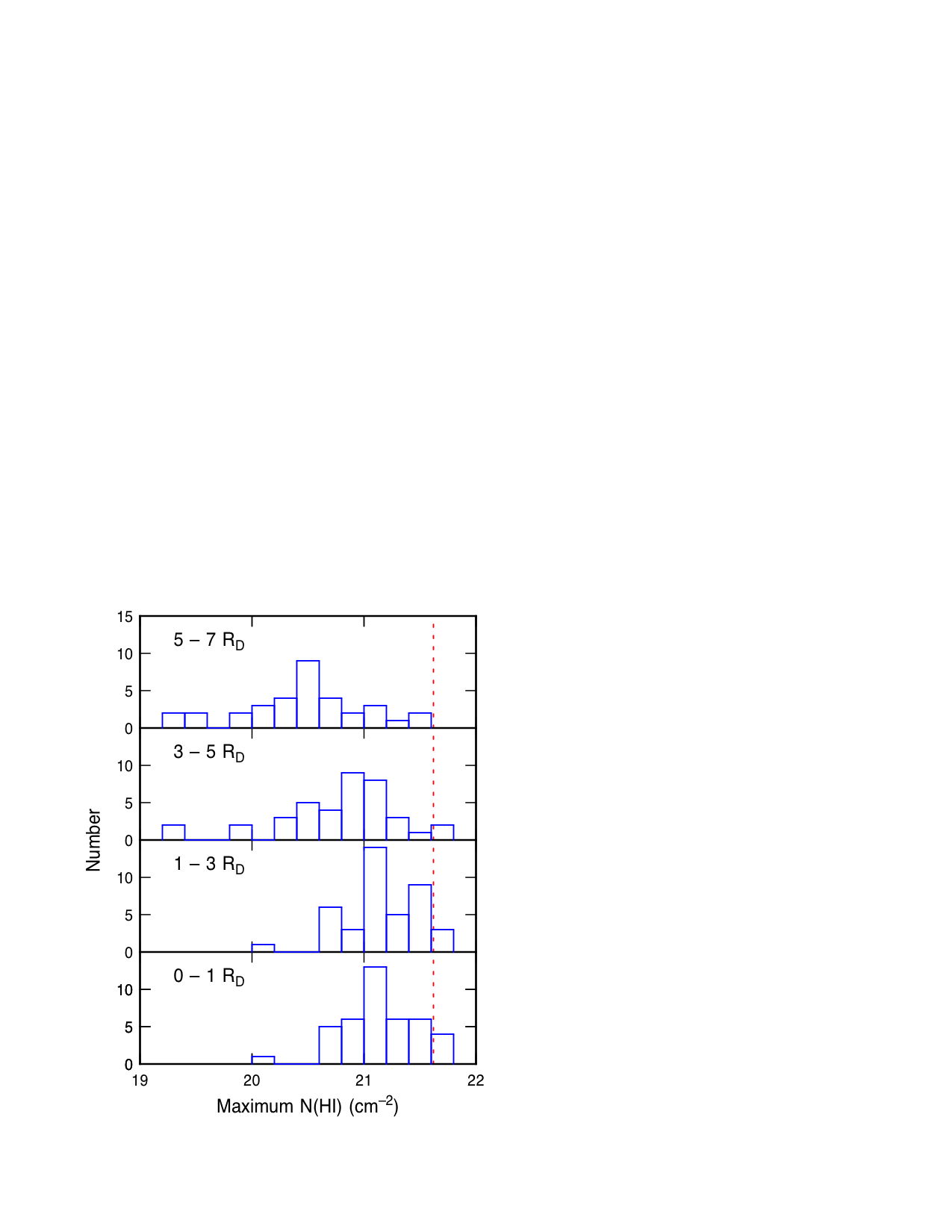}
\vskip -0.75truein
\caption{Histograms of the maximum value of \nhi\ in four radial intervals for all of the
galaxies in this survey. The vertical dashed line is the \HI\ column density where the opacity
to FUV radiation is unity at a metallicity of 1/8 solar, which is typical for these dIrrs. Most
values of maximum \nhi\ are much less than this opacity limit, which suggests that molecule
formation is in the weak field limit, occurring where the H$_2$ formation rate at
a cloud surface balances the photo dissociation rate.
\label{fig-maxnhi}}
\end{figure}

Figure \ref{fig-maxnhi} shows histograms of the absolute maximum values of \nhi\ for all
galaxies in the four radial bins. The distributions of \nhi\ shift toward lower values
with increasing radius, and never exceed a maximum value (for our spatial resolution) of
$\log N(HI)\sim21.8$.  These trends and maximum values give insight into the undetected
high-density phases of the interstellar medium.

Molecule formation requires shielding of the molecules from background FUV
radiation, and this shielding results from a combination of dust extinction and line
absorption by similar molecules in the cloud envelopes \citep{krum09,bialy16}. According
to \cite{bialy16}, the total column density of \HI\ through a slab that turns molecular
with increasing depth depends only on the parameter $\alpha G$, which, for $H_2$
formation on dust grains, is given by their equation (22),
\begin{equation}
\alpha G = 0.59 I_{\rm FUV}\left({{100\;{\rm cm}^{-3}}\over{n_{\rm cloud}}}\right)\left({{9.9}\over
{1+8.9{\tilde\sigma_g}}}\right)^{0.37}.
\label{eq:aG}
\end{equation}
Here, $\alpha$ is the ratio of the H$_2$ photodissociation rate in free space to the
H$_2$ formation rate, $G$ is the ratio of the dust to gas absorption cross sections times
the maximum bandwidth in Hz for H$_2$ absorption in the Lyman-Werner bands, $I_{\rm FUV}$
is the local radiation field in units of the value in the solar neighborhood
\citep{draine78}, $n_{\rm cloud}$ is the cloud density, and $\tilde\sigma_g$ is the
normalized metallicity-dependent dust absorption cross-section per hydrogen nucleon. The
product $\alpha G$ is the ratio of the shielded photo-dissociation rate to the H$_2$
formation rate. When $\alpha G$ is small, the \HI\ layer at the edge of a cloud is limited
by molecular self-absorption and photo-dissociation; when $\alpha G$ is large, dust
opacity determines the \HI\ column density at the cloud edge; $\alpha G=1$ represents the
borderline between dust and dissociation attenuation for incoming radiation in the \HI\
layer around an H$_2$ cloud.

For the dIrrs in our sample, the average metallicity is
$Z^\prime\sim1/8$ relative to solar, so the non-normalized dust absorption
cross-section per hydrogen nucleon is $\sigma_{\rm
g}\sim1.9\times10^{-21}Z^\prime\sim2.4\times10^{-22}$ cm$^{-2}$. The dust opacity
in an \HI\ cloud, $\sigma_{\rm g} N(HI)$, equals 1 when $N(HI)$ is the inverse of
$\sigma_{\rm g}$, or $4.2\times10^{21}$ cm$^{-2}$, which corresponds to $42\;M_\odot$
pc$^{-2}$ including He and heavy elements. This column density for opaque gas is
much higher than our average \nhi, and close to the upper limit to \nhi\ in the
inner regions. Figure \ref{fig-maxnhi} has a dotted red line at this value, which
is $\log N(HI)=21.62$. The histograms reach this value only in the inner parts of
the galaxies.  Thus, most of the local \HI\ peaks in the main parts and outer disks
of our galaxies are not optically thick from dust even though there is star
formation at these radii. This implies that much of the H$_2$ gas associated with
star formation, perhaps all but the molecules that are deeply embedded in dense
self-gravitating cores, is self-shielded rather than dust-shielded. Then $\alpha G$
is small and $\sigma_{\rm g} N(HI)\sim\alpha G$ \citep[from Eq. 19 in the limit of
small $\alpha G$ in][considering that $N(HI)$ in that equation is for half of the
cloud]{bialy16} for $N(HI)$ at the threshold of molecule formation. Small $\alpha
G$ is also likely for our galaxies because $I_{\rm FUV}$ in equation (\ref{eq:aG})
is small; these are low surface brightness galaxies with star formation surface
densities that are lower than that in the solar neighborhood by a factor of 10 to
100.


Thus, molecule formation in our sample of dIrrs is mostly in the weak-field limit
\citep{bialy16}, where the shielding column around a cloud is given by a balance between
the integrated $H_2$ formation rate and the integrated $H_2$ destruction rate in the
incident radiation field. Then $N(HI)_{\rm max}n_{\rm cloud}\propto I_{\rm FUV}$ and dust
is relatively unimportant. Since $I_{\rm FUV}$ varies with radius, as well as from galaxy to galaxy,
this is presumably why the peak \nhi\ is not a fixed absolute
value for different annuli in Figure \ref{fig-pdfrd}.

We previously determined the average radial variations of $N(HI)$ and $I_{\rm FUV}$ for
20 LITTLE THINGS dIrrs to be $\propto\exp(-\xi r/R_{\rm D})$ for relative scale length
$\xi=0.45\pm0.32$ and $0.98\pm0.80$, respectively, where $R_{\rm D}$ is the $V$-band disk
scale length \citep{bruce15b}. The FUV intensity comes from the ratio of the volume
emissivity to the opacity, which is the same as the ratio of the surface brightness of
FUV to the surface density of gas. If the peak \nhi\ is proportional to the average
$<N(HI)>$, as suggested for many of the galaxies by Figure \ref{fig-pdfrd}, then the
average $H_2$ cloud density in the shielding envelope varies with galactocentric radius
as the ratio of $I_{\rm FUV}$ to $<N(HI)>$, which is $n_{\rm cloud}\propto\exp(-\xi_{\rm
cloud}r/R_{\rm D})$ for $\xi_{\rm cloud}=0.53\pm1.07$ \citep{bruce15b}. We can compare
this density variation to the expected pressure variation considering that the average
pressure scales with the square of the column density, which for these gas-dominated and
\HI-dominated regions, is just $N(HI)$. Thus, the ISM pressure falls with radius as
$P_{\rm ISM}\propto N(HI)^2\propto\exp(-[0.9\pm0.64]r/R_{\rm D})$. This pressure
variation is similar to that of the product of the $H_2$ transition cloud density and the
square of the \HI\ velocity dispersion (for which $\xi_{\rm disp}=0.10\pm0.10$), which is
$P_{\rm cloud}=\exp(-[0.73\pm1.5]r/R_{\rm D})$. It follows that the envelopes of the
clouds with a fixed maximum relative \nhi\ are probably at the weak field limit of
molecule formation and in approximate pressure equilibrium with the ambient ISM, as also found in \cite{bruce15b}.

The PDFs  inside and outside \rbr, seen in Figure \ref{fig-pdfrbr}, are similar in
shape to those in Figure \ref{fig-pdfrd}, especially the PDFs inside the breaks.
Generally the PDFs beyond \rbr\ appear smoother and broader than the outer PDFs in
Figure \ref{fig-pdfrd} because the area beyond \rbr\ is larger. Thus, although the
stellar mass surface density shows a change at \rbr, the \HI\ column
density PDF does not.

\subsection{Skewness and kurtosis}

To determine the third and fourth moments of the \nhi\ distributions, following
\citet{smc10}, we normalized the \nhi\ by subtracting the average \nhi\ and dividing by
the standard deviation (see their Equation 4). The normalization was determined over the
region that the skewness and kurtosis was being measured: the entire galaxy (\S
\ref{sec-skwide}), within an annulus (\S \ref{sec-skann}), or in individual cells over
the galaxy (\S \ref{sec-skkernel}). 
The skewness and kurtosis values 
for the distribution functions of \HI\ column density were measured from those functions plotted on a linear-linear scale 
(rather than log-log as in Figures 2 and 3), using pixels from the integrated \HI\ maps. 
This is the procedure
given by Equations 7 and 8 in \citet{smc10}. 
The uncertainties are $\sqrt{6/N}$ and
$\sqrt{24/N}$ for skewness and kurtosis, respectively, where $N$ is the number of \HI\
map beams included in the measurements (area divided by beam area).

\subsubsection{Galaxy-wide values} \label{sec-skwide}

\begin{figure}
\epsscale{0.85}
\plotone{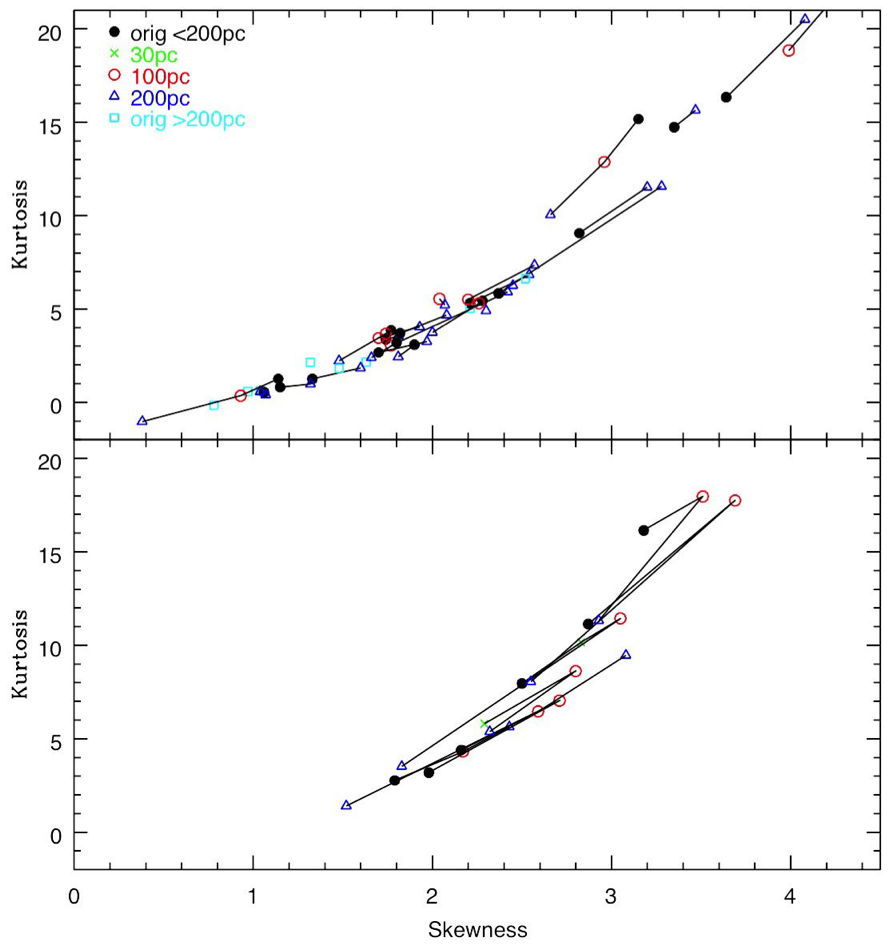}
\caption{
Kurtosis versus skewness measured over the entire \HI\ moment 0 maps of the galaxies.
Pixels with \nhi\ $\le 15\times10^{18}$ cm$^{-2}$ were eliminated to ensure good S/N.
Values are plotted for the original \HI\ maps, with separate symbols for beam-sizes that have major axes that are
$<$200 pc in size (``orig $<$200pc'') and those with beam-sizes that are $>$200 pc (``orig $>$200 pc'').
Values are also plotted for maps smoothed to 30 pc, 100 pc, and 200 pc beam-sizes, where possible.
Original beam-sizes that are close to one of the smoothed values are plotted with the smoothed symbol and color.
The solid black lines connect different maps of the same galaxy in order to facilitate comparison
of values at different map spatial resolutions.
NGC 1569 at 200 pc resolution, with a kurtosis value of 38, is not plotted.
Bottom panel shows those galaxies where
skewness and kurtosis values do not decline as the beam-size increases: DDO 69,
DDO 75, DDO 187, DDO 210, IC 10, UGC 8508, WLM.
\label{fig-skwhole}}
\end{figure}

We have measured the skewness and kurtosis values for the entire \HI\ distributions of
each galaxy. Pixels with \nhi\ $\le 15\times10^{18}$ cm$^{-2}$ were eliminated to ensure
good S/N. Values were measured on the original integrated \HI\ maps and on maps smoothed
to beam-sizes of 30 pc, 100 pc, and 200 pc, where possible. Values for all maps are given
in Table \ref{tab-results} and plotted in Figure \ref{fig-skwhole} where different maps
for the same galaxy are connected with a solid black line. We see that, for a given
galaxy, generally skewness and kurtosis values decline as the beam-size increases. This
makes sense: as the beam-size increases peaks and valleys in the \HI\ column density
distributions get smoothed out and so the deviations from the average decrease in
amplitude. However, there are some exceptions where kurtosis values increase as beam-size
increases, and in some cases, then decrease for the highest beam size of 200 pc: DDO 69,
DDO 75, DDO 187, DDO 210, IC 10, UGC 8508, WLM.
As a whole, the results for beam-sizes
$>$200 pc fall in the left hand corner of Figure \ref{fig-skwhole} (skewness $< 2.6$ and
kurtosis $< 7$), but results for beam-sizes $\sim$200 pc and $\sim$100 pc are found
throughout the full range of values. Values for the two galaxies smoothed to 30-pc
beam-sizes fall in the middle of the range shown in Figure \ref{fig-skwhole}. Here we
will compare galaxies with comparable beam-sizes.

\begin{figure}
\epsscale{0.5}
\plotone{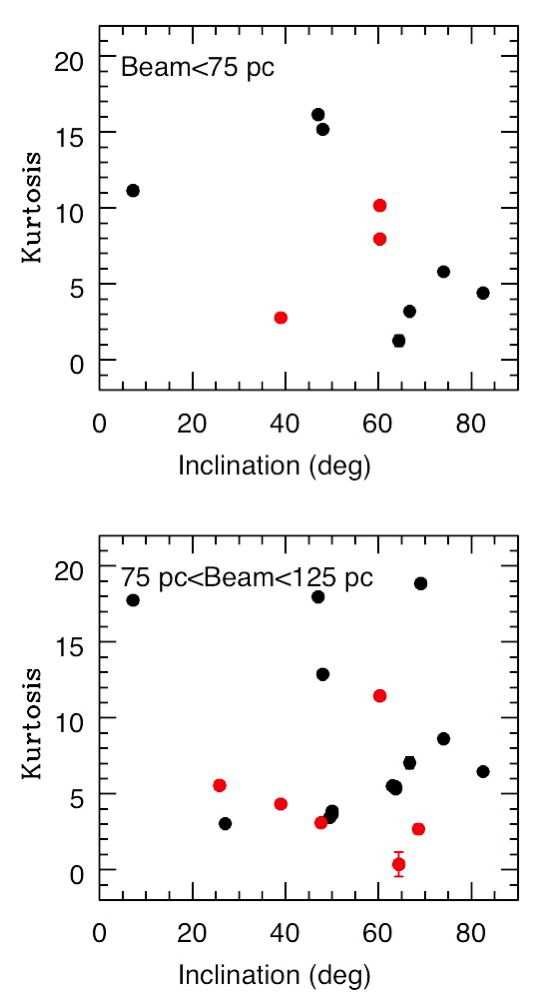}
\caption{
Kurtosis measured over the entire \HI\ moment 0 maps of the galaxies plotted against inclination of the disk.
The inclinations are given by \citet{oh15} for galaxies marked with an ``H'' in Figure \ref{fig-pdfrd} (black points)
and by \citet{he06} for galaxies marked with an ``O'' in Figure \ref{fig-pdfrd} (red points).
Pixels with \nhi\ $\le 15\times10^{18}$ cm$^{-2}$ were eliminated to ensure good S/N.
Maps with beam-sizes $<$75 pc are plotted in the upper panel, and maps with beam-sizes
75 pc to 125 pc are plotted in the lower panel. Maps with beam-sizes $>$125 pc are not plotted here.
We do not see a significant effect of inclination, and its effect on angular resolution, on kurtosis values.
\label{fig-incl}}
\end{figure}

Another factor that could potentially affect spatial resolution is the inclination of the
galaxy. The more inclined the disk is, the more material a given sight-line is
integrating over and so the effective spatial resolution is lower. In Figure
\ref{fig-incl} we plot the galaxy-wide kurtosis values against inclination of the \HI\
disk. We have grouped the galaxies into beam-sizes: $<$75 pc and 75 pc to 125 pc.
Galaxies with beam-sizes larger than 125 pc are not included in this plot.
Note that black points were determined using inclinations derived from the \HI\ kinematics
and red points from the optical morphology. We differentiate them in order to show that there is
no trend with type of inclination determination.
If inclination
is a significant factor, we would expect kurtosis values to be higher for more face-on
systems. In the upper panel, where the maps with beam-sizes $<$75 pc are plotted, we see
such a trend in the upper envelope of points.
However, there is a large scatter in
kurtosis values for inclinations $\le$60\arcdeg. In the bottom panel, where maps with
beam-sizes 75 pc to 125 pc are plotted, we see a general scatter of kurtosis values with
no trend. From these plots, we conclude that inclination could be a factor in the highest
resolution maps, but that this does not affect our results any more than comparing
galaxies over our range in resolutions.

\begin{figure}
\epsscale{0.9}
\vskip -5truein
\plotone{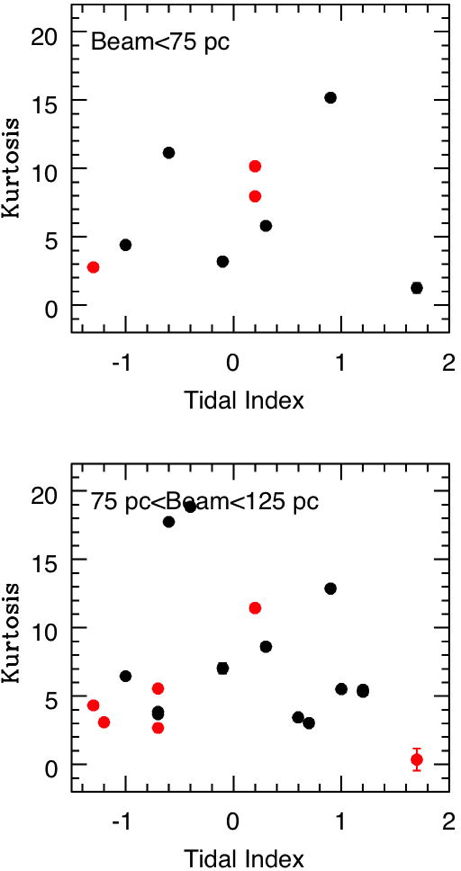}
\caption{
Kurtosis measured over the entire \HI\ moment 0 maps of the galaxies plotted against the tidal index of the galaxy
from \citet{tidalindex} as given by \citet{zhang12}.
The larger the tidal index, the stronger the gravitational disturbance exerted by neighboring galaxies.
Nearly isolated galaxies have zero or negative values.
Black points are galaxies marked with an ``H'' in Figure \ref{fig-pdfrd} and red points are
galaxies marked with an ``O'' in Figure \ref{fig-pdfrd}.
Pixels with \nhi\ $\le 15\times10^{18}$ cm$^{-2}$ were eliminated to ensure good S/N.
Maps with beam-sizes $<$75 pc are plotted in the upper panel, and maps with beam-sizes
75 pc to 125 pc are plotted in the lower panel. Maps with beam-sizes $>$125 pc are not plotted here.
We do not see a correlation of kurtosis with tidal index.
\label{fig-tidalindex}
}
\end{figure}

One difference between the LITTLE THINGS dwarfs and the SMC is their relative isolation.
The LITTLE THINGS dwarfs were chosen to be relatively far from other galaxies, especially
giant systems, and not (obviously) engaged in an interaction with another galaxy. On the other hand,
the SMC is clearly interacting with the LMC and the Milky Way.
\citet{smc10} suggest that high skewness and kurtosis values could result from
large-scale gravitational effects in a galaxy-galaxy interaction.
The tidal index, computed by \citet{tidalindex} and collected by \citet{zhang12} for the LITTLE THINGS
sample, is a measure of the strength of gravitational disturbance
exerted by neighboring galaxies. Nearly isolated galaxies have zero or negative values,
and the LITTLE THINGS sample has tidal indices ranging from -1.3 to 1.7.
In Figure \ref{fig-tidalindex} we plot
kurtosis measured over the entire \HI\ moment 0 maps of the galaxies against the tidal index of the galaxy.
We do not see a correlation of galaxy-wide kurtosis with tidal index.

\begin{figure}
\epsscale{0.4}
\plotone{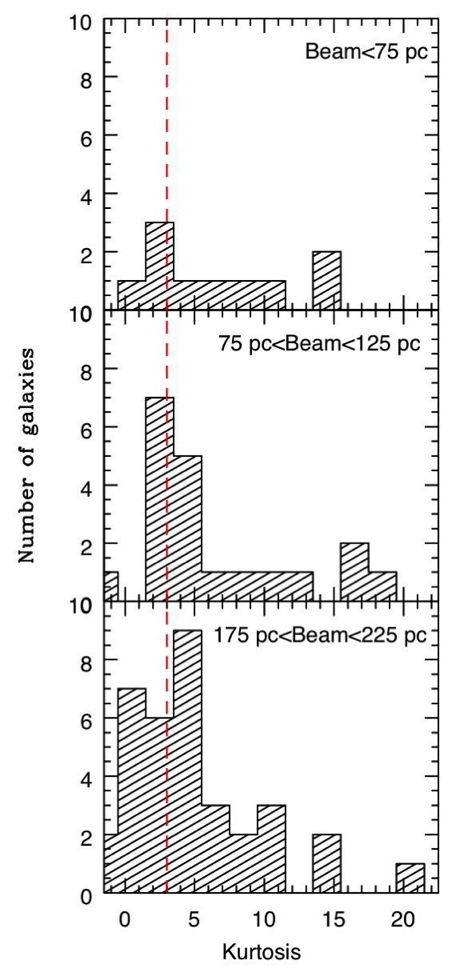}
\caption{
Number of galaxies with certain kurtosis values measured over the entire \HI\ moment 0 maps of the galaxies.
Pixels with \nhi\ $\le 15\times10^{18}$ cm$^{-2}$ were eliminated to ensure good S/N.
The maps are grouped by beam-size: beam-size $<$75 pc, beam-size between 75 pc and 125 pc,
and beam-size between 175 pc and 225 pc.
A galaxy in the top panel will also appear in the middle and bottom panels, and a galaxy that appears in the
middle panel will also appear in the bottom panel.
NGC 1569 at 200 pc resolution, with a kurtosis value of 38, is not plotted.
The vertical red, dashed line marks a kurtosis value of 3; above this value the gas is supersonic according
to the models of \citet{smc10}.
Most (60\%) of the galaxies have kurtosis values $\le5$ regardless of angular resolution of the map,
but numbers range up to 19.
\label{fig-histkurt}}
\end{figure}

Figure \ref{fig-histkurt} shows a histogram of kurtosis values with maps grouped by
beam-size: $<$75 pc,  between 75 pc and 125 pc, between 175 pc and 225 pc. We see that
typical values of kurtosis are $\le5$, and 60\% of the sample has values in this range.
These values are comparable to the kurtosis value
of 2.5 measured by \citet{smc10} for the SMC and values measured in spiral galaxies
\citep{spirals}.
The rest of the systems have kurtosis values that range from 5 up to 19.

\begin{figure}
\epsscale{0.7}
\vskip -3truein
\plotone{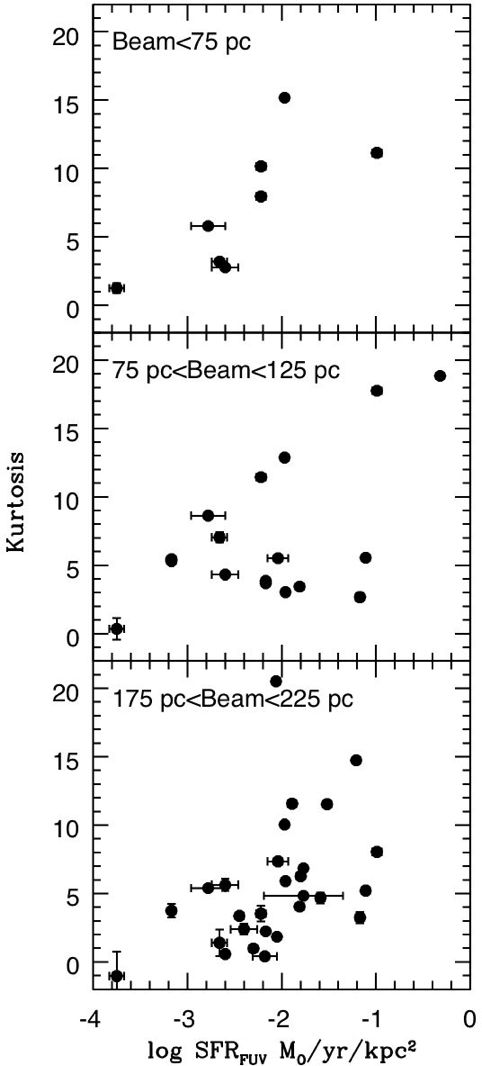}
\caption{
Kurtosis values measured over the entire \HI\ moment 0 maps of the galaxies versus integrated SFR
determined from the FUV \citep{hunter10}.
Uncertainties are shown in both quantities but in many cases the uncertainties are smaller than the point symbol.
Pixels with \nhi\ $\le 15\times10^{18}$ cm$^{-2}$ were eliminated to ensure good S/N.
The maps are grouped by beam-size: beam-size $<$75 pc, beam-size between 75 pc and 125 pc,
and beam-size between 175 pc and 225 pc.
A galaxy in the top panel will also appear in the middle and bottom panels, and a galaxy that appears in the
middle panel will also appear in the bottom panel.
NGC 1569 at 200 pc resolution, with a kurtosis value of 38, is not plotted.
There is a general trend of higher kurtosis values for higher SFR.
\label{fig-kurtsfr}}
\end{figure}

Figure \ref{fig-kurtsfr} shows kurtosis values plotted against integrated galactic star
formation rates (SFRs) determined from the FUV \citep{hunter10}. Again, maps are grouped
by beam-size, which determines the spatial resolution:  $<$75 pc,  between 75 pc and 125
pc, between 175 pc and 225 pc. Although there is a lot of scatter, there is a general
trend of higher SFRs being accompanied by higher kurtosis values at all three spatial
resolutions. We expect higher turbulence to be associated with the energy input from star
forming regions \citep[for example, in the ionized gas,][]{moiseev15}, and we expect
highly turbulent regions to produce more star formation in compressed shells and other
structures \citep{con3}.

\subsubsection{Radial variations} \label{sec-skann}

To look at how kurtosis might vary with radius in galaxies, we have selected a subsample
of galaxy maps that were chosen to have at least 20 beam-sizes along the semi-major axis.
This ensures that annuli can be constructed in which there are enough independent beams
to determine a statistically significant kurtosis value. The radii were chosen to be 6,
8.5, 10.5, 12.5, 14, 15.5, etc.\ times the beam major axis. This gives at least 100 beam
areas per annulus, assuming every pixel has emission. To maximize resolution, we use the
original unsmoothed \HI\ maps, except for IC 10 and IC 1613 for which we use the 30-pc
beam maps. We also eliminated pixels with values less than $15\times10^{18}$ cm$^{-2}$
for S/N, and eliminated radii where the number of pixels is such that the uncertainty in
kurtosis exceeds 1. The uncertainty is based on the number of beam areas included in the
pixels with emission.

Radial plots of kurtosis profiles and FUV surface photometry $\mu_{FUV}$ are shown
in Figure \ref{fig-kurtann}. We see a variety of kurtosis profiles. In three
galaxies (DDO 70, DDO 210, DDO 216) the kurtosis values are predominantly flat with
radius and in 7 (DDO 50, DDO 155, DDO 187, IC 10, M81dwA, NGC 4163, WLM) there is a
gradual increase of kurtosis with radius. In 5 galaxies (DDO 53, DDO 63, DDO 75,
NGC 2366, NGC 4214) there are particularly high values in the outer disk and in
three others (DDO 69, IC 1613, NGC 1569) there are spikes in kurtosis closer in to
the center of the galaxy. We also see that kurtosis and $\mu_{FUV}$, which is
proportional to the recent star formation activity, are not related in the
azimuthal-averaged annuli. This is in spite of the fact that we found a general
trend of higher SFR with kurtosis for galaxy-wide integrated values in Figure
\ref{fig-kurtsfr} for galaxies with beam-sizes less than 75 pc. The
trend of increasing kurtosis with radius is partly the result of a change in the
ratio of the maximum value of $N(HI)$ in each annulus divided by the value at the
peak of the PDF, as discussed in Section \ref{sec-discuss}.

We explore the possible effects of nearby neighbors in Figure \ref{fig-slope}, where we plot
the slope of a linear fit to kurtosis as a function of radius against the tidal index of
\citet{tidalindex}.
We have eliminated spikes in kurtosis for the fits.
The fits are shown as green lines in Figure \ref{fig-kurtann}.
Nearly isolated galaxies have zero or negative tidal indices, so most of the galaxies
would be considered isolated. The one known interacting system in these plots, NGC 1569, is
plotted in red and it has a tidal index of -0.4, so it too would be considered isolated.
We do find a general trend in the sense that a higher tidal index
means a more constant kurtosis value with radius.

In Figure \ref{fig-annkurtvsskew} we plot kurtosis versus skewness for each annulus.
For 72\% of the sample of 18, we see a strong correlation between skewness and kurtosis,
with kurtosis values ranging from about $-1$ to 17 and skewness values ranging from about 0 to 4.
The exceptions are DDO 70, DDO 155, DDO 210, DDO 216, and M81dwA,
where the annuli clump around kurtosis and skewness values of 0-2.

\subsubsection{Kernel analysis} \label{sec-skkernel}

We selected a sub-sample of \HI\ maps for kernel analysis. In particular, we chose galaxies with beam sizes of order 100 pc or less
in order to have high angular resolution, but included all maps --- original beam and smoothed beams --- of the galaxy up to a beam-size
of 100 pc to allow comparison of resolution effects.
We also eliminated galaxies that were too small to contain a reasonable number of kernels.
The final sub-sample is given in Table \ref{tab-kernels}.
For these galaxies, we divided the \HI\ maps into a grid of adjacent, contiguous kernels or cells
\citep[for more details, see][]{spirals}. Each kernel is 32 of the \HI\ map pixels $\times$ 32 pixels, where
the \HI\ map pixel is 1.5\arcsec$\times$1.5\arcsec\ except in DDO 216 where the pixel size is 3.5\arcsec$\times$3.5\arcsec.
Typically the original beam-size is several \HI\ map pixels across.
The equivalent of a 32-pixel length in parsecs is given in Table \ref{tab-kernels} for each galaxy.
We normalized the column densities
and calculated skewness and kurtosis values of the \HI\ distributions within each kernel.
Then we constructed maps of skewness and kurtosis in which the 32 pixel $\times$ 32 pixel area of the original map is
replaced by a single kurtosis or skewness value.
We geometrically converted FUV images to match the scale and orientation of the \HI\ maps,
measured the FUV magnitude in each kernel, and constructed an FUV map for comparison to the \HI\ higher order moments.
These maps are shown in Figure \ref{fig-kernelmaps}.
Each square in these images is one kernel.
For each map of each galaxy, we also plotted skewness versus kurtosis for the ensemble of kernels, one point per kernel.
These plots are shown in Figure \ref{fig-kernelkurtvsskew}.

For DDO 69, IC 1613, and WLM, in Figure \ref{fig-kernelmaps} we show the maps produced from
30 pc and 100 pc beam-size maps. For the most part, the maps are the same, regardless of resolution.
The exception is the skewness map of WLM that shows a difference in skewness values between
the inner and outer galaxy in the 30-pc map that is not seen in the 100-pc map.
The rest of the galaxy maps are based on the 100-pc beam-size maps.
We see that for most of the galaxies, the skewness and kurtosis values are fairly uniform across the galaxy.
However, sometimes (DDO 50, DDO 69, DDO 70, IC1613, WLM, NGC 2366, NGC 4214)
there is a general increase in skewness values around the edges of the galaxy, which
we also found for spiral galaxies \citep{spirals}.
Also as we saw in spiral galaxies, most often there is not a noticeable correlation of kurtosis or skewness values with FUV features.
However, DDO 50, IC 1613, and WLM do show a small correlation.

In the skewness versus kurtosis plots (Figure \ref{fig-kernelkurtvsskew}),
a comparison of different beam-sizes for a given galaxy
(DDO 69, DDO 70, DDO 75, IC 1613, WLM; DDO 187 does not have enough points)
shows that a map with lower spatial resolution tends to show a poorer correlation of
skewness with kurtosis, 
while the higher resolution map tends show more of a correlation.
The exceptions are DDO 70 and IC 1613, which show a poor correlation  
at both resolutions.
Presumably the higher resolution beams are a more accurate
indication of the motions of the gas in the galaxy.
Of the rest of the galaxies (excluding DDO 216 for the limited number of points),
DDO 50, DDO 53, DDO 155, DDO 210, IC 10, NGC 1569, NGC 2366, and NGC 4214
show a strong correlation between kurtosis and skewness
with most values ranging from -1 to 8 for kurtosis and -0.5 to 3 for skewness.
A few kernels have kurtosis values as high as 30.
In all, of the 13 galaxies in the kernel sample 
69\% show a correlation between kurtosis and skewness among all the kernels.
Of the 8 galaxies showing a strong correlation between kurtosis and skewness among the kernels, one did not show
such a pattern in the annuli (DDO 210) and a second did not have enough points for a determination (DDO 155, Figure \ref{fig-annkurtvsskew}).
Of the 5 galaxies showing  weak correlations among the kernels, one also did not show a correlation
among annuli (DDO 70).
Thus, for the most part, the annuli results and the kernel results are similar with respect to correlations
between kurtosis and skewness.

\section{Discussion} \label{sec-discuss}

The PDF of interstellar column density is close to log-normal in the solar
neighborhood \citep{lombardi10}, as expected for compressive turbulence
\citep{vazquez94,pvs98,pnj97,lemaster08,federrath10}.  A log-normal is a Gaussian
distribution in the logarithm of the column density. When the ISM column densities
are replotted on a linear scale, there is a fat tail at high \nhi\ which contains
the same information about turbulence as the original log-normal.
This fat tail is shown in Figure \ref{fig-logloglinear_turbulence7}. The top two
panels have log-normal distribution functions, evaluated as discussed below, with 6
different widths and normalized by area. On the left these functions are plotted in
log coordinates, and on the right the same functions are plotted in linear
coordinates. The tail at high density on the linear plot gets larger relative to
the peak of the distribution function as the log-normal width increases. The bottom
two panels show distribution functions with 6 widths and convolved with power law
tails, simulating turbulence in self-gravitating clouds (Eq. \ref{eq:pdftotal}
below).

In a physical model of turbulence, the tail is caused by numerous
tiny compressed regions. For an intrinsically log-normal
function, the tail results in skewness and kurtosis in the linear distribution.
Skewness and kurtosis tend to increase with Mach number because the
dispersion in a log-normal PDF increases with Mach number \citep{pnj97}.
They also increase when the maximum value of $N(HI)$ in the
distribution function increases relative to the value at the peak of the function
because that extends the measured part of the fat tail further.


We model the dependence of kurtosis on skewness to reproduce Figures
\ref{fig-skwhole}, \ref{fig-annkurtvsskew}, and \ref{fig-kernelkurtvsskew} by
constructing log-normal distribution functions and measuring their skewness and
kurtosis on a distribution with linear coordinates as we did for the
observations.  We consider two types of PDF functions, written in
terms of spatial density; they are considered to be approximately the same shape
for column density. One is a pure log-normal as shown on the top of
Figure \ref{fig-logloglinear_turbulence7},
\begin{equation}P(\ln \rho)d\ln \rho=(2\pi D^2)^{-1/2}e^{-0.5\left(\ln(
\rho/\rho_{\rm pk})/D\right)^2}
\end{equation}
where the density at the peak, $\rho_{\rm pk}$, is related to the average density by
\begin{equation}
\rho_{\rm pk}=\rho_{\rm ave}e^{-0.5D^2},\end{equation} and the dispersion, D, is related to
the Mach number, M, by \citep{pnj97}
\begin{equation}
D^2=\ln(1+0.25{\cal M}^2).\label{eq:D}\end{equation}

The second PDF is a log-normal modified by a power law at high density, which is
appropriate for self-gravitating clouds \citep{kritsuk10}. We use the formalism in
\cite{elmegreen11} where the final PDF is the convolution of a power law radial
density profile in a self-gravitating cloud with a log-normal distribution locally
inside the cloud for the average density at that position. The Mach number is
assumed to be constant inside the cloud. This gives
\cite[see][]{elmegreen11}
\begin{equation}
P_{\rm PDF,total}(y)= {{3{\cal
C}}\over{\alpha(2\pi)^{0.5}}}\int_{1/{\cal C}}^{1}
\exp\left(-{{\ln^2(yze^{0.5D^2})}\over{2D^2}}\right) {{\left(z{\cal
C}-1\right)^{(3-\alpha)/\alpha}}\over{D\left({\cal
C}-1\right)^{3/\alpha}}}dz,\label{eq:pdftotal}
\end{equation}
per unit $\ln y$, where $y=\rho/\rho_{\rm edge}$ is the local normalized density,
including turbulent fluctuations, and $z=\rho_{\rm edge}/\rho_{\rm ave}(r)$ is the
inverse of the average density, not including the turbulent fluctuations, both
normalized to the density at the edge of the cloud, $\rho_{\rm
edge}$. The average density is assumed to vary with radius as a
cored power-law to simulate the effects of self-gravity,
\begin{equation}
\rho_{\rm ave}(r)=\rho_{\rm edge} {{ r_{\rm edge}^\alpha + r_{\rm
core}^\alpha}\over {r^\alpha + r_{\rm
core}^\alpha}}.\label{eq:rho}
\end{equation}
The core radius, $r_{\rm core}$, avoids a density singularity. The degree of central
condensation,
\begin{equation}{\cal C}={{\rho_{\rm ave}(r=0)}\over{\rho_{\rm edge}}},
\end{equation}
is varied to simulate how strongly self-gravitating the cloud is.

The kurtosis and skewness of a distribution function with a fat tail
depend sensitively on the upper limit for $N(HI)$. We modify these PDFs to
include upper and lower density cutoffs by converting the density in the equation
to a density of observation using
\begin{equation}
\rho_{\rm obs}={{1}\over{1/\rho+1/\rho_{\rm upper}}}
\end{equation}
if $\rho>\rho_{\rm ave}$, and
\begin{equation}
\rho_{\rm obs}=\rho+\rho_{\rm lower}
\end{equation}
if $\rho<\rho_{\rm ave}$, and where $\rho_{\rm lower}=0.01$ and $\rho_{\rm
upper}=10$ and 100 in two cases. At large $\rho$ the observed density
converges to the constant value $\rho_{\rm upper}$, which is interpreted as the
limiting density before molecules form. Higher density gas shows this limiting
density from \HI\ in the shielding layer. Below $\rho_{\rm lower}$, the \HI\ is assumed
to disappear because of ionization and heating in a hot medium. Phase
transitions like this were also necessary to fully characterize the pdf of \HI\
column density in the LMC \citep{elmegreen01}.

The bottom left-hand side of Figure \ref{fig-kurtosismodel_turbulence5b} shows the
density PDFs in the non-gravitating case for Mach numbers of 0.52, 0.85, 1.40,
2.29, 3.76, and 6.17 as a sequence of increasing width, and for
$\rho_{\rm upper}=10$ (red) and 100 (blue). The kurtosis versus skewness is shown
in the top-left, using the same colors. Kurtosis and skewness both
increase with Mach number at first and then decrease again for the largest Mach
number. They have the same relationship to each other as in Figures
\ref{fig-skwhole}, \ref{fig-annkurtvsskew}, and \ref{fig-kernelkurtvsskew},
keeping in mind that the models are for density and the observations are of column
density. The observations span a range in PDF values that is a factor of
$10^2$-$10^3$ on the vertical axis (Fig. \ref{fig-pdfrd}), so the distinction
between the $\rho_{\rm upper}=10$ and 100 models is not clear in Figure
\ref{fig-kurtosismodel_turbulence5b} (we can rule out $\rho_{\rm
upper}=1000$ because it is too broad).

The right-hand side of Figure \ref{fig-kurtosismodel_turbulence5b} shows density
PDFs, kurtosis, and skewness for the self-gravitating model with center-to-edge
contrast ${\cal C}=100$, using the same Mach numbers and lower and upper limits as
on the left. The self-gravitating model is also a good match to the
observations, although in this case, the kurtosis and skew decrease with increasing
Mach number for $\rho_{\rm upper}=100$.

The change in kurtosis with Mach number is a result of a change in
the PDF close to the peak. For the non-gravitating case, the PDF is symmetric in
log-log coordinates, so increasing Mach number simply increases the strength of the
tail, and with it the kurtosis. In the gravitating case, the power-law part from
the radial gradient of density in the cloud is more prominent at low Mach number
because then turbulence does not interfere as much with the power law. This is the
most asymmetric case near the peak because the power law at high density makes the
tail stronger relative to the lower density part. At higher Mach number the
turbulence dominates the power law and the PDF becomes more curved, as in the
non-gravitating case.  Then the kurtosis decreases with increasing Mach number.

\begin{figure}
\epsscale{0.9}
\plotone{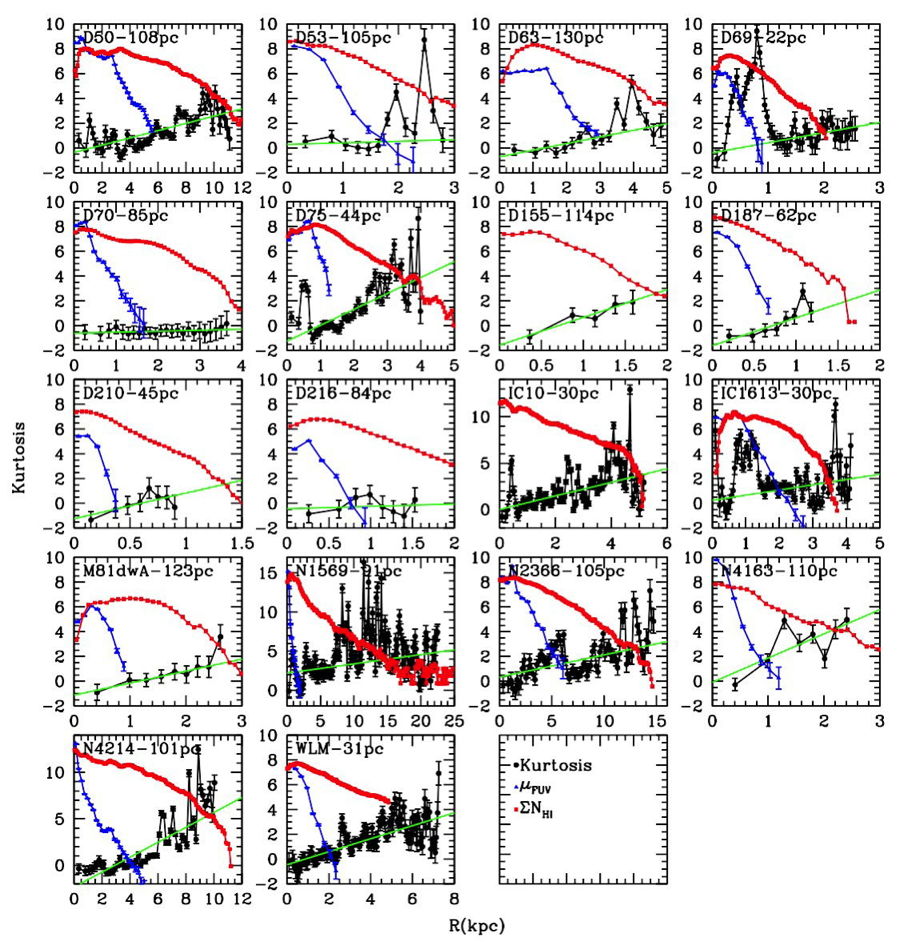}
\caption{
Kurtosis values measured in elliptical annuli versus the radii of the midpoints of the annuli.
Pixels with \nhi\ $\le 15\times10^{18}$ cm$^{-2}$ were eliminated to ensure good S/N.
Only galaxies with maps in which there are at least 20 beam-sizes along the semi-major axis were included.
The average of the major and minor axes of the beam size are given next to the galaxy name in each panel.
The ellipse parameters (center, inclination and minor-to-major axis ratio, position angle of the major axis)
are given by \citet{oh15} for galaxies marked with an ``H'' in Figure \ref{fig-pdfrd}
and by \citet{he06} for galaxies marked with an ``O'' in Figure \ref{fig-pdfrd}.
We see no correlation between features in the kurtosis profiles and those in the FUV profiles.
Green lines are linear fits to the kurtosis profiles, not including spikes.
\label{fig-kurtann}}
\end{figure}

\begin{figure}
\epsscale{0.9}
\vskip -5truein
\plotone{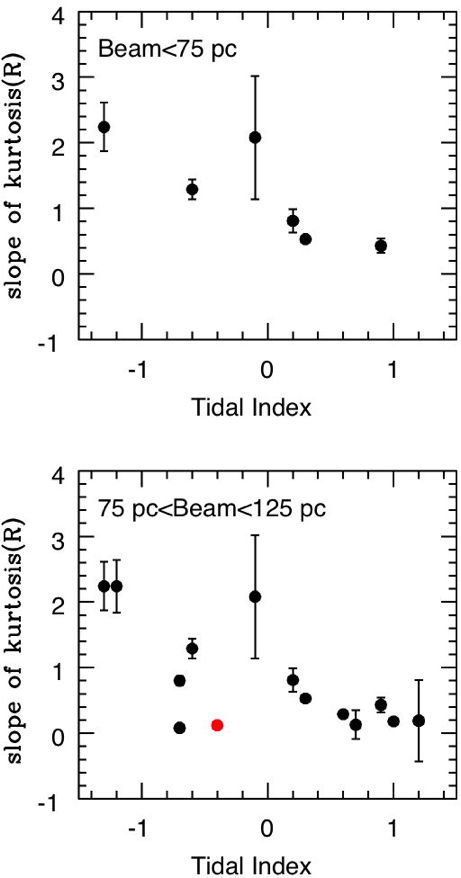}
\caption{
Slope of a linear fit to kurtosis radial profiles (Figure \ref{fig-kurtann}) against the tidal index of
\citet{tidalindex}.
We have eliminated the spikes in kurtosis from the fits.
The fits are shown in green in Figure \ref{fig-kurtann}.
Nearly isolated galaxies have zero or negative tidal indices.
The one known interacting system in these plots, NGC 1569, is plotted in red.
We find a general trend in the sense that a higher tidal index
generally means a more constant kurtosis value with radius.
\label{fig-slope}}
\end{figure}

\begin{figure}
\epsscale{0.9}
\plotone{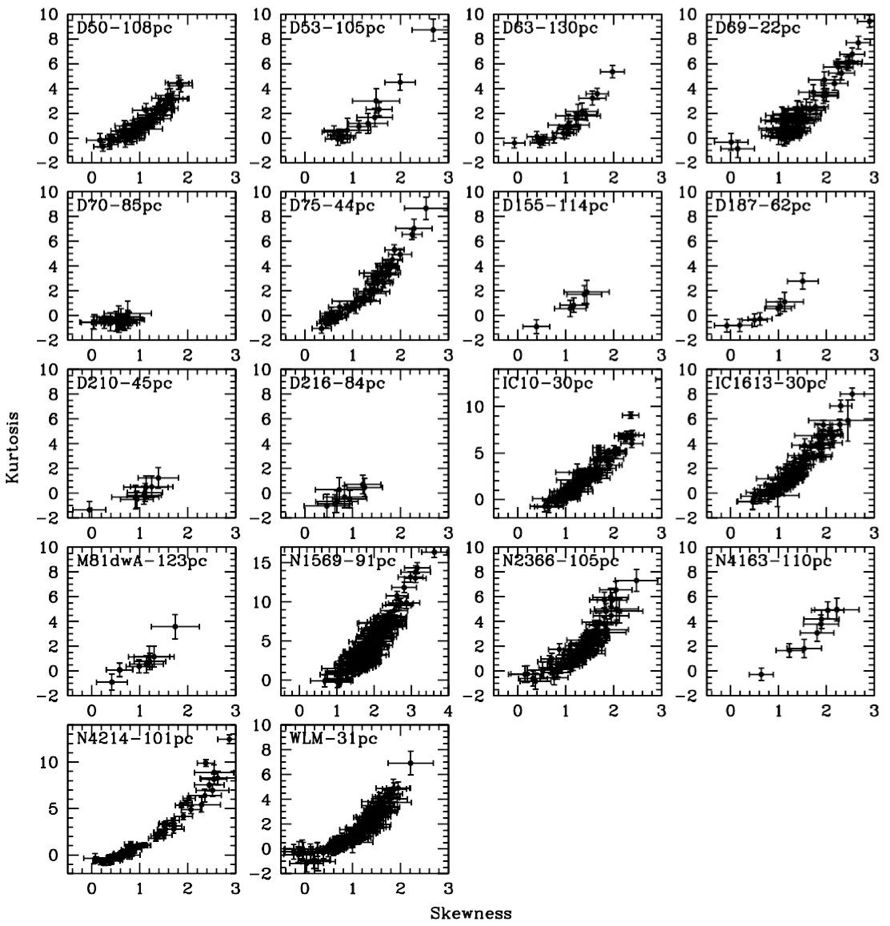}
\caption{
Kurtosis values plotted against skewness values measured in elliptical annuli.
Pixels with \nhi\ $\le 15\times10^{18}$ cm$^{-2}$ were eliminated to ensure good S/N.
Only galaxies with maps in which there are at least 20 beam-sizes along the semi-major axis were included.
The average of the major and minor axes of the beam size are given next to the galaxy name in each panel.
The ellipse parameters (center, inclination and minor-to-major axis ratio, position angle of the major axis)
are given by \citet{oh15} for galaxies marked with an ``H'' in Figure \ref{fig-pdfrd}
and by \citet{he06} for galaxies marked with an ``O'' in Figure \ref{fig-pdfrd}.
Most (72\%) of the 18 galaxies show a strong correlation between kurtosis and skewness values.
\label{fig-annkurtvsskew}}
\end{figure}

\begin{figure}
\epsscale{0.9}
\plotone{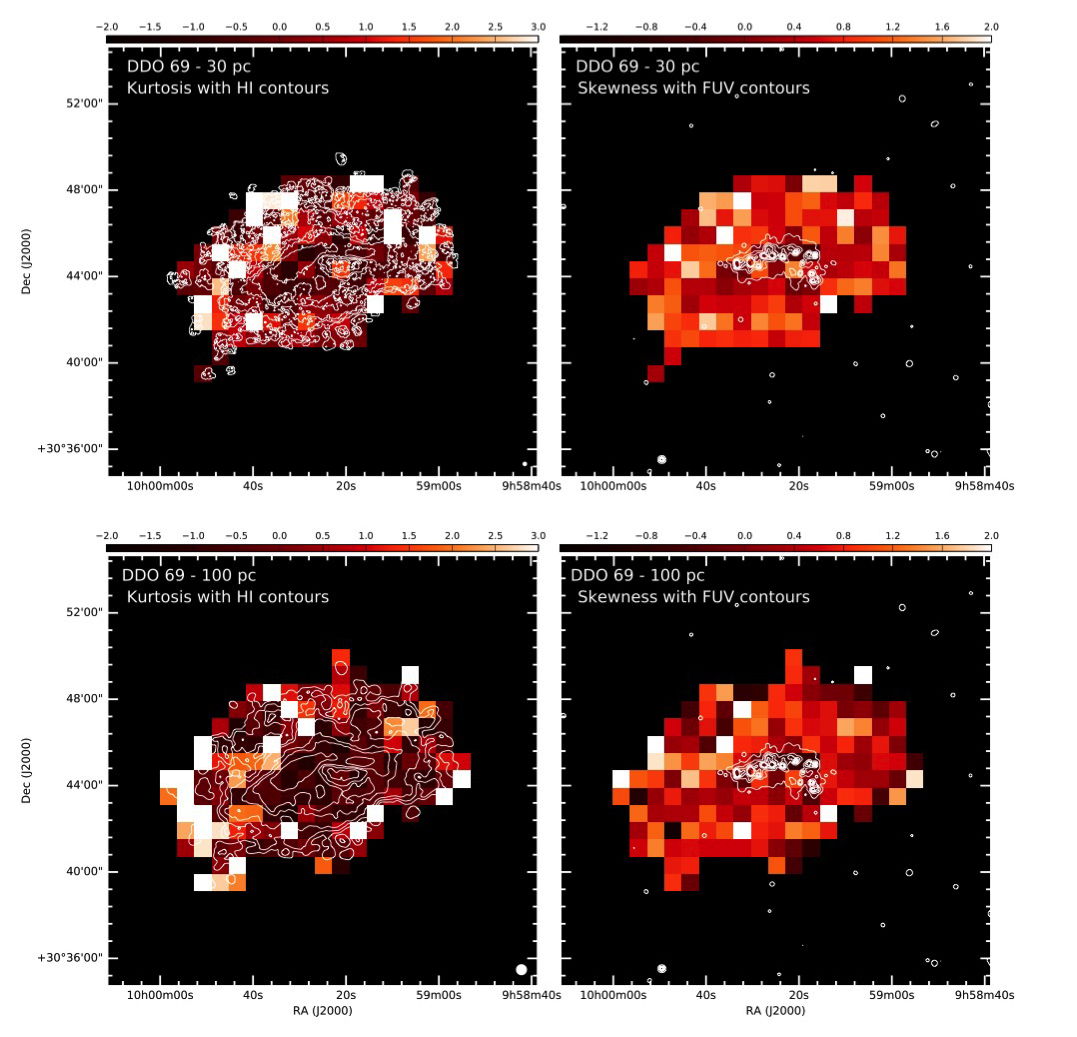}
\caption{Maps of kurtosis (left) with contours of integrated \HI\ and of
skewness (right) with contours of FUV. The kurtosis and skewness maps are determined
in contiguous kernels or cells that are 32 of the \HI\ map pixels $\times$ 32 pixels in size, so each square in these
images is one kernel.
The size of the kernel in parsecs is given in Table \ref{tab-kernels}.
\HI\ map beam-size is indicated as a solid white ellipse in the right corner of the left image, and is given
in parsecs next to the galaxy name in the upper left.
The \HI\ contours are 5, 30, 100, 300, 500, 1000, and $3000\times10^{18}$ atoms cm$^{-2}$,
except for DDO 50, IC 1613, NGC 1569, and WLM for which the 30 and $300\times10^{18}$ atoms cm$^{-2}$
contours are left out.
The kurtosis and skewness maps show little correlation with star formation as seen in the FUV.
\label{fig-kernelmaps}}
\end{figure}

\begin{figure}
\epsscale{0.8}
\plotone{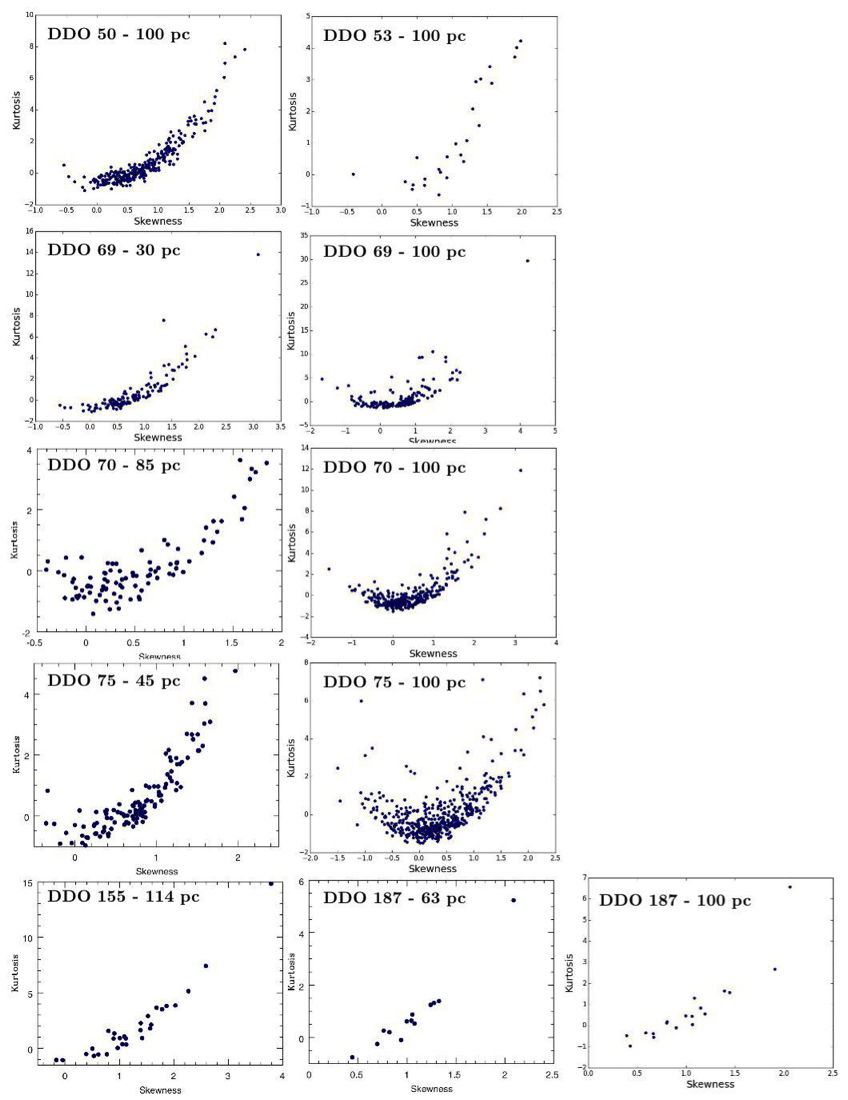}
\caption{Kurtosis values versus skewness values for 32 pixel $\times$ 32 pixel kernels.
The beam size of the \HI\ map is given next to the galaxy name in each plot.
For maps with the original beam, the given beam size is the average of the major
and minor axes of the beam. For smoothed maps, the minor and major axes are the same.
We assume that the higher angular resolution
maps of these galaxies are a better indication of gas turbulence.
Many of the galaxies
show a strong correlation between kurtosis and skewness among the kernels.
\label{fig-kernelkurtvsskew}}
\end{figure}

\begin{figure}
\epsscale{0.8}
\plotone{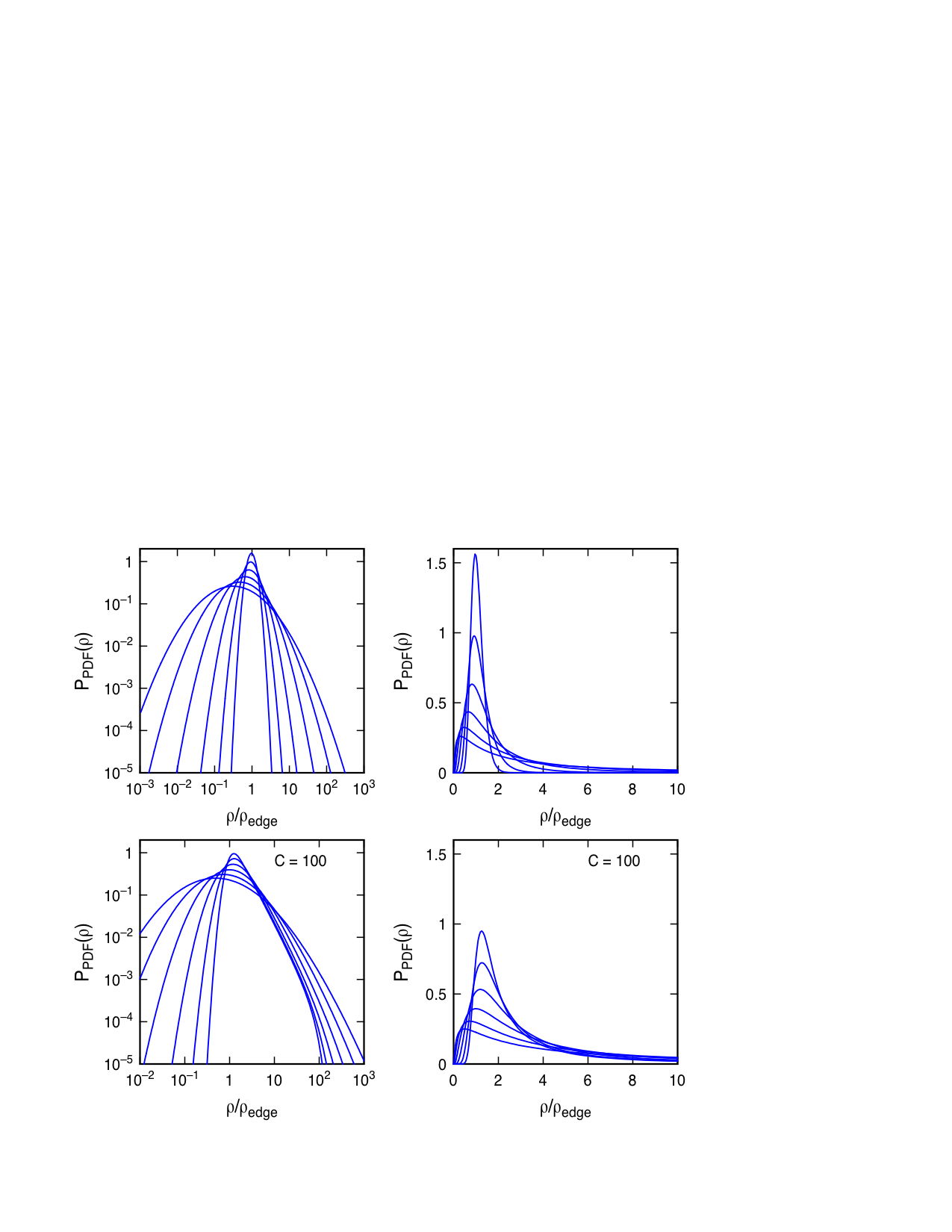}
\vskip -0.75truein
\caption{{\it Top}:  Density PDFs in the
pure log-normal case from equations (2)-(4)
for Mach numbers of 0.52, 0.85, 1.40, 2.29, 3.76, and 6.17 as a sequence
of increasing width. On the left, the PDFs are plotted in log-log coordinates and on the
right they are plotted in linear coordinates. The tail responsible for kurtosis and
skew is always present, and it gets relatively larger for wider log-normal functions.
{\it Bottom}: PDFs for the same Mach numbers, now using equation (\ref{eq:pdftotal}) with
a ratio of average-to-edge density, ${\cal C}$, equal to 100. For small Mach numbers, the
power-law density profile dominates the PDF and the high-density side becomes nearly a power law.
For large Mach numbers, the log-normal still dominates in the high-density side and the PDF
remains curved in log-log coordinates. On the right, the same distributions are plotted
in linear coordinates.
\label{fig-logloglinear_turbulence7}}
\end{figure}

\begin{figure}
\epsscale{0.95}
\plotone{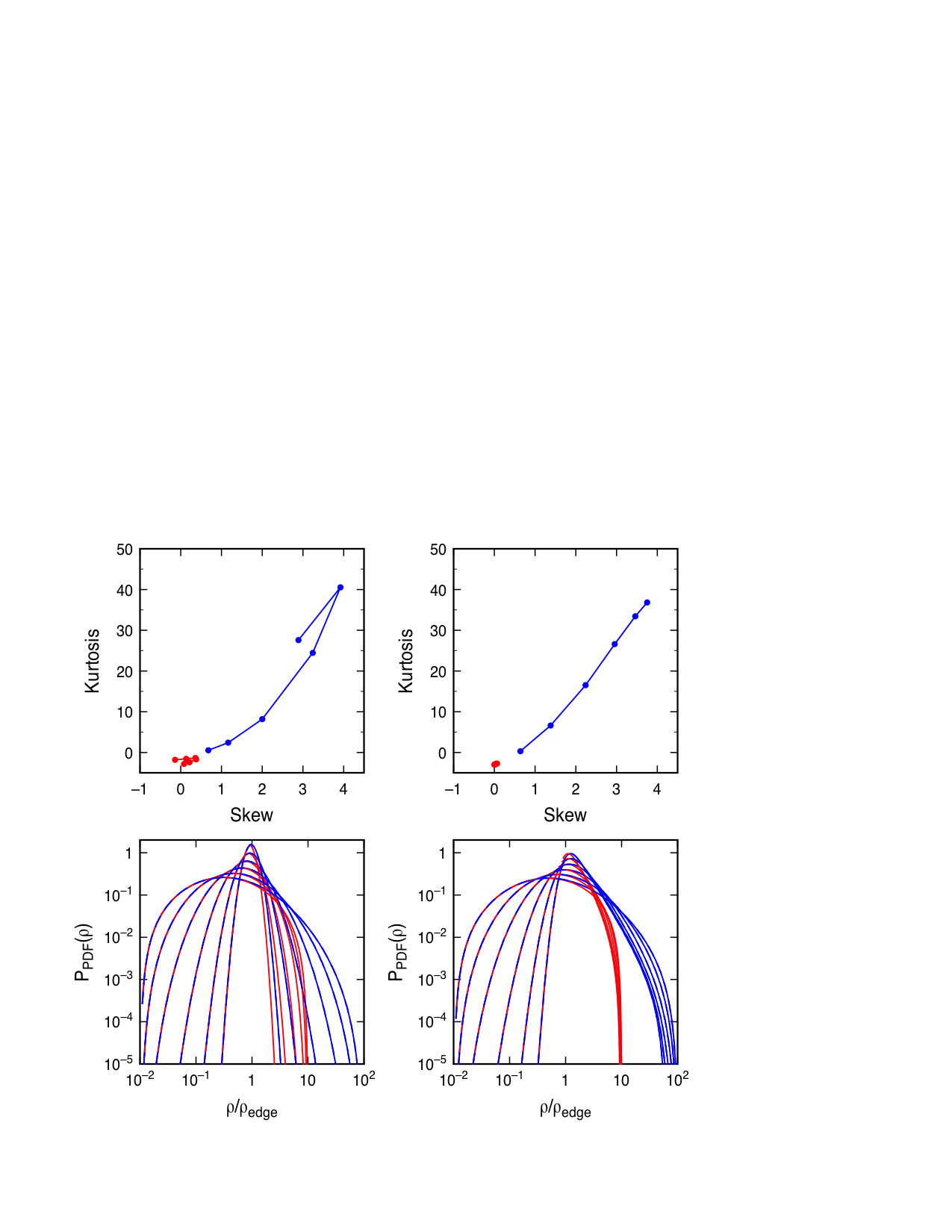}
\vskip -1.0truein
\caption{Kurtosis model.
{\it Bottom left}:  Density PDFs in the
pure log-normal case for Mach numbers of 0.52, 0.85, 1.40, 2.29, 3.76, and 6.17 as a sequence
of increasing width, and for $\rho_{\rm upper}=10$ (red), 100 (blue). The curves
are dashed blue and red for low $\rho$ because the functions are the same for the two
values of $\rho_{\rm upper}$.
{\it Top left}: Kurtosis versus the skewness, using the same colors. Kurtosis
and skewness both increase with Mach number, and they have the same relationship as in
Figures \ref{fig-skwhole}, \ref{fig-annkurtvsskew}, and \ref{fig-kernelkurtvsskew},
keeping in mind that the models are for density and the observations are of column density.
{\it Right}: Density PDFs, kurtosis, and
skewness for the self-gravitating model with center-to-edge contrast ${\cal C}=100$, using
the same Mach numbers and upper limits as on the left.
\label{fig-kurtosismodel_turbulence5b}}
\end{figure}

\begin{figure}
\epsscale{0.95}
\plotone{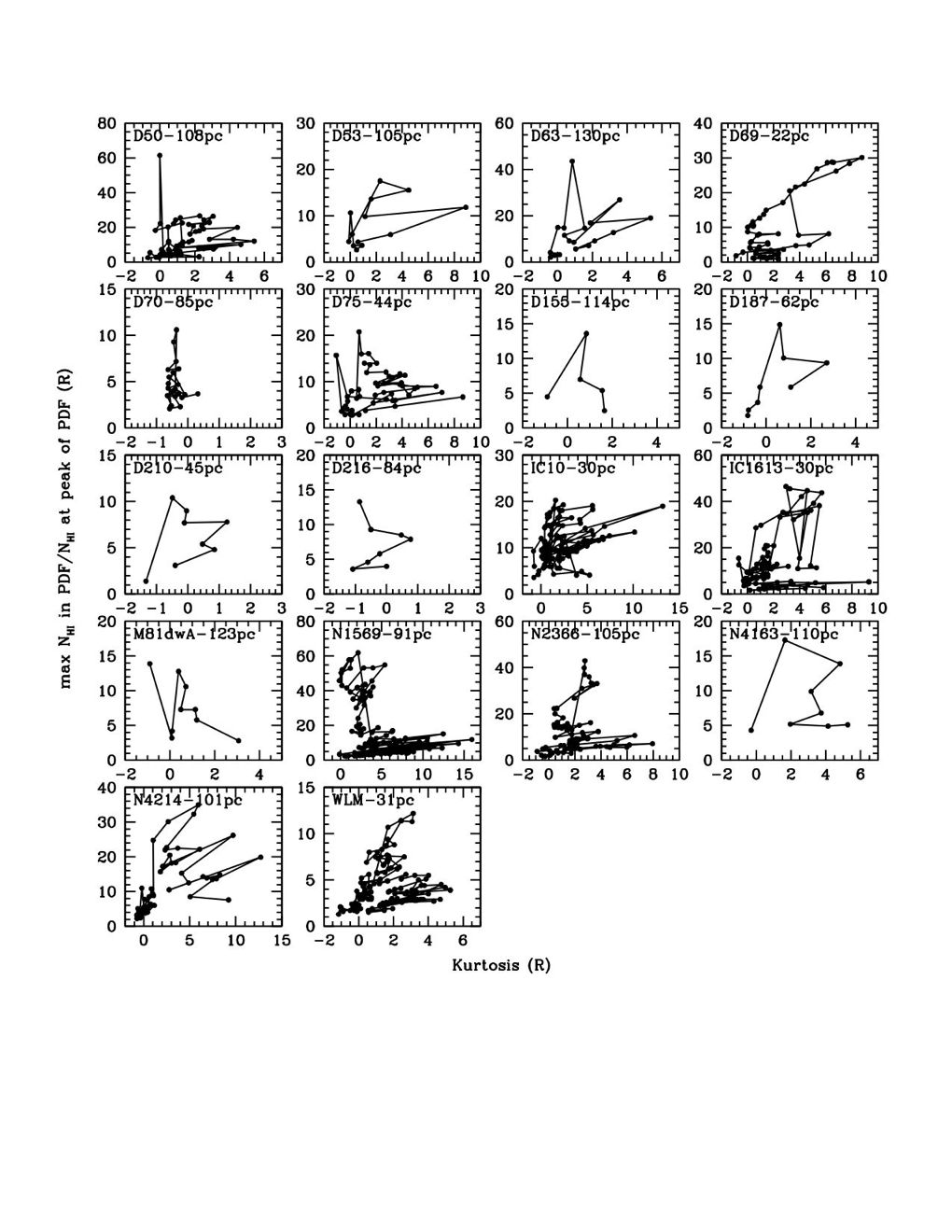}
\vskip -1.75truein
\caption{The ratio of the maximum value of $N(HI)$ in each annulus to the value at the peak
of the PDF in that annulus, versus kurtosis for the same annulus. Each point is a different
annulus and the points are connected by line segments for increasing radius.  Excursions
toward high ratios of column density often correspond to high values of kurtosis, indicating
that kurtosis depends partly on how far out the \HI\ can be observed in the PDF tail
before it disappears or converts to molecules.
\label{fig-ratpeak}}
\end{figure}

Figure \ref{fig-kurtosismodel_turbulence5b} suggests that the range of kurtosis and
skewness observed for dIrr galaxies is the result of variations in the Mach number
combined with a variable upper limit to the column density from the
conversion of atomic to molecule gas. The effect of the upper limit to $N(HI)$ may
dominate the radial variation in the kurtosis shown in Figure \ref{fig-kurtann}. We
plotted (not shown) the ratio of the maximum value of $N(HI)$ to the value at the
peak of the PDF in each radial annulus versus the radius of the annulus. The radial
distribution of rising or flat values for this ratio resembled the trends of
kurtosis in Figure \ref{fig-kurtann}.  This implies that there is a correlation
between kurtosis and the observational cutoff in the fat tail. Figure
\ref{fig-ratpeak} shows this correlation more directly for each galaxy in Figure
\ref{fig-kurtann}. The interconnected points in the figure are the values for each
annulus traced over increasing radius. There is a tendency for simultaneous
excursions on each axis, although the slopes of these excursions vary from galaxy
to galaxy and sometimes have two values inside any given galaxy.

\section{Summary} \label{sec-summary}

We have presented PDFs and higher order moment (skewness and kurtosis) analysis of the
integrated \HI\ column density distributions in the LITTLE THINGS sample of nearby dIrr
galaxies. This analysis follows that of \citet{smc10} for the SMC. We find that
galaxy-wide values of kurtosis are similar to that found for the SMC for about 60\% of our sample,
with values ranging as high as 19.
The kurtosis values for whole galaxy
PDFs roughly correlate with integrated SFRs such that higher SFRs correspond more often
to higher kurtosis values.
Kurtosis and skewness values calculated for annuli within a
subsample of 18 galaxies with beam-sizes less than or equal to 130 pc most often (72\%) show
a correlation between kurtosis and skewness values with no correlation with the FUV radial profile.
We also analyzed a grid of 32 pixel $\times$ 32 pixel kernels (160 pc $\times$ 160 pc to 840 pc $\times$ 840 pc areas) in the 30-pc and
100-pc maps of a subsample of 13 galaxies. There we also found a correlation
of kurtosis with skewness for many of the galaxies with most kurtosis values ranging from $-1$ to 8,
especially at
the highest resolution, where small \HI\ peaks at high density
could be resolved. However, there was no 
correlation with local FUV features.

These results are consistent with a model where the typical dIrr galaxy has a
turbulent multi-phase ISM with subsonic to weakly supersonic motions that are excited on
scales larger than the local star-forming regions but still related to the integrated
star formation rate. Higher spatial resolution shows more \HI\ substructure and suggests
higher Mach numbers.

The distribution functions for the maximum values of \nhi\ in four radial intervals
show a decreasing trend with radius that suggests \HI\ disappears at high density
by conversion to molecules at a column density determined by the balance between
H$_2$ formation on grain surfaces and H$_2$ destruction by ambient radiation. Aside
from the innermost radii in a few galaxies, the \HI\ column of gas
rarely becomes optically thick to UV radiation from dust extinction,
but converts to H$_2$ because of self-shielding when it is still
optically thin. As a result, molecule formation is sensitive to local pressure and
radiation. We find that the shielding envelopes of the H$_2$ clouds
could be in pressure equilibrium with the general interstellar medium. Because all
radial intervals have star formation and therefore dense gas, and essentially all
of the \HI\ column densities peak in optically thin gas, there should be a
substantial amount of diffuse H$_2$ in these galaxies, with conversion to CO
possibly limited to the dense cores where stars actually form.

Models of log-normal density PDFs or log-normals convoluted with a
power law to simulate self-gravitating clouds were used to explain the observed
trends between kurtosis and skewness as a result of variations in the PDF width,
which is related to Mach number and the upper HI cutoff where
presumably molecules form. Kurtosis and skewness in linear PDFs for \nhi\ are
expected because of the conversion in plotting coordinates from log-log to
linear-linear.

\acknowledgments

DAH is grateful to the Lowell Observatory Research Fund for funding for this research,
and to John and Meg Menke for a donation to Lowell Observatory that covered part
of the page charges.
GH and EM appreciate funding from the National Science Foundation grant AST-1461200 to Northern Arizona University
for Research Experiences for Undergraduates summer internships and Dr.\ Kathy Eastwood and
Dr.\ David Trilling for running the REU program.
We also appreciate suggestions on the manuscript from an anonymous referee.

Facilities: \facility{VLA} \facility{GALEX}

\clearpage

\clearpage

\plotone{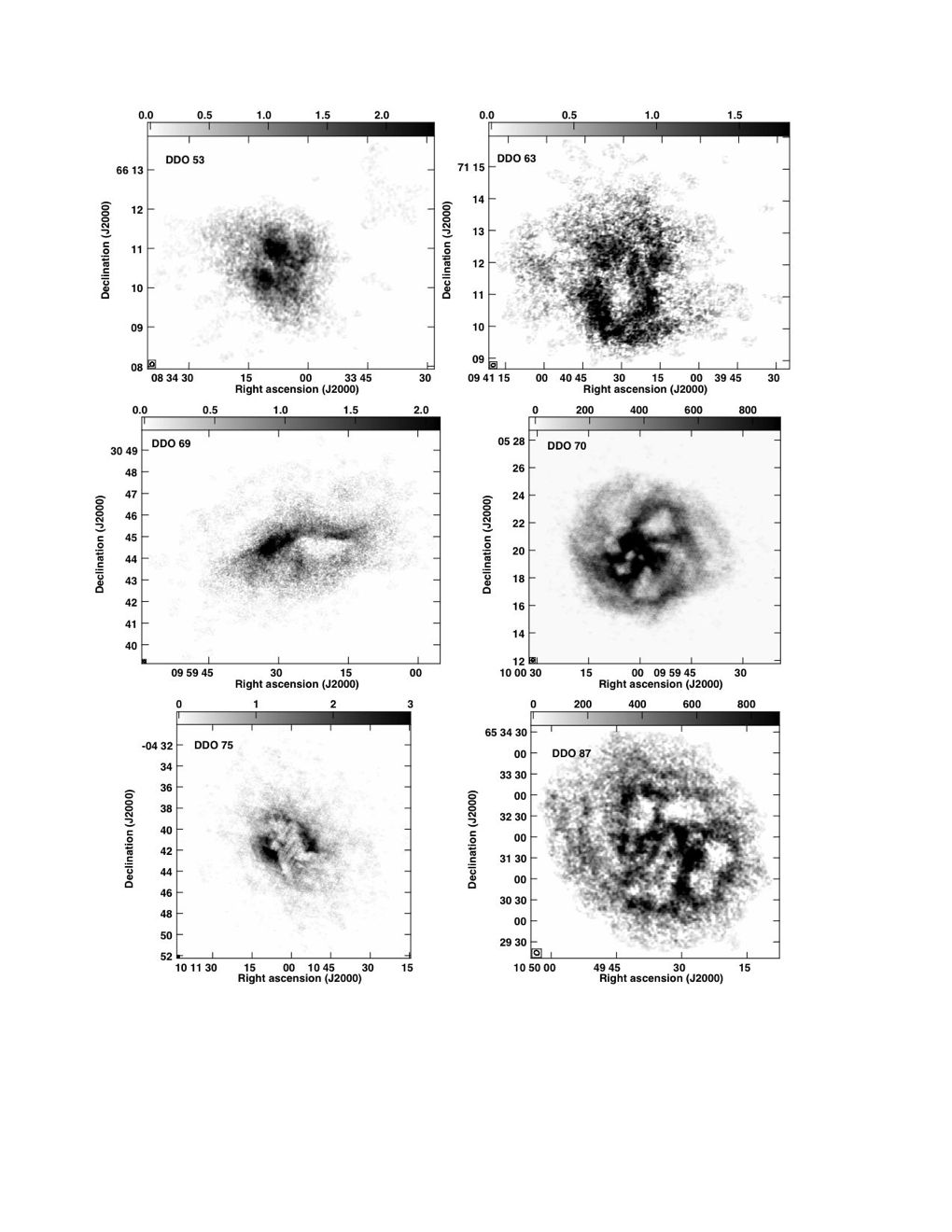}

\clearpage

\plotone{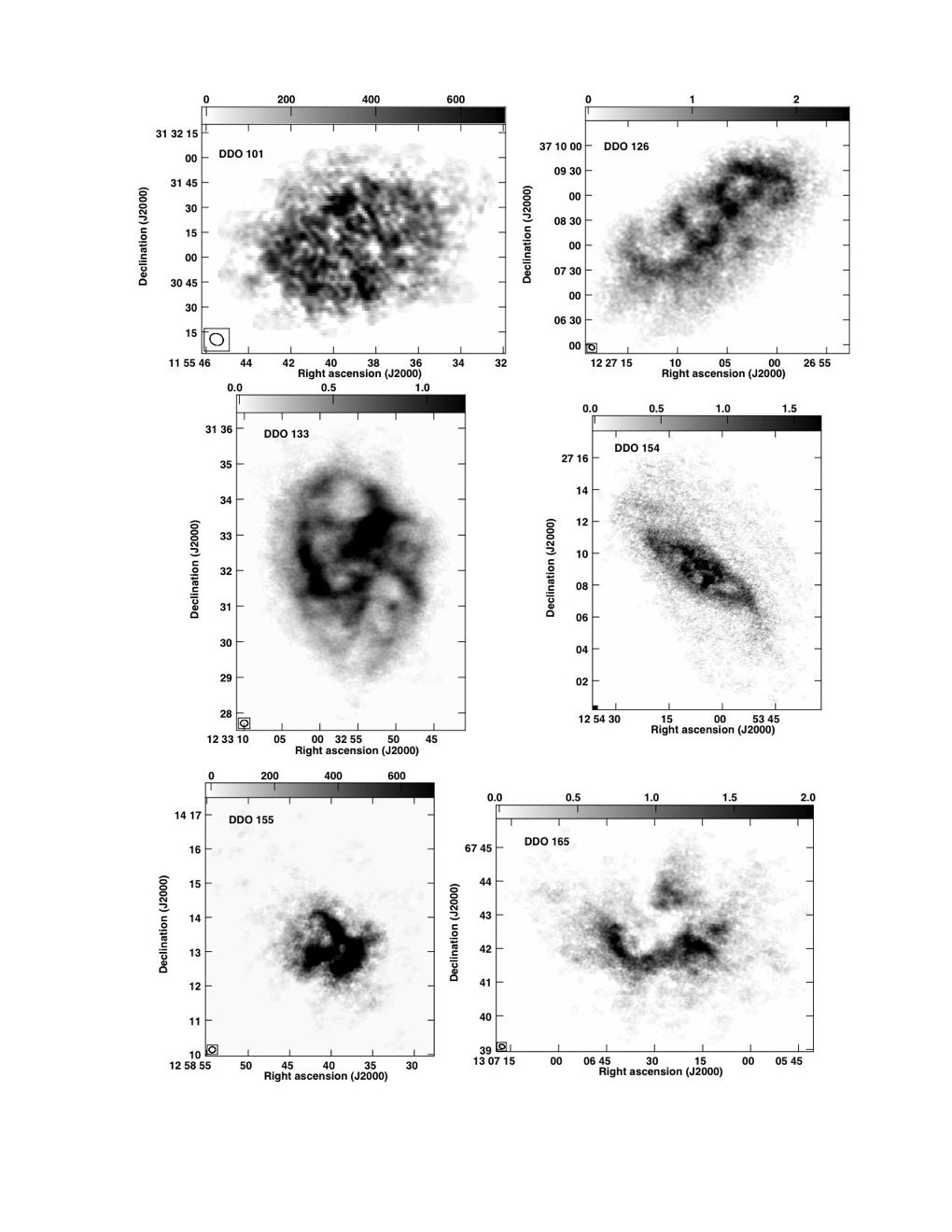}

\clearpage

\plotone{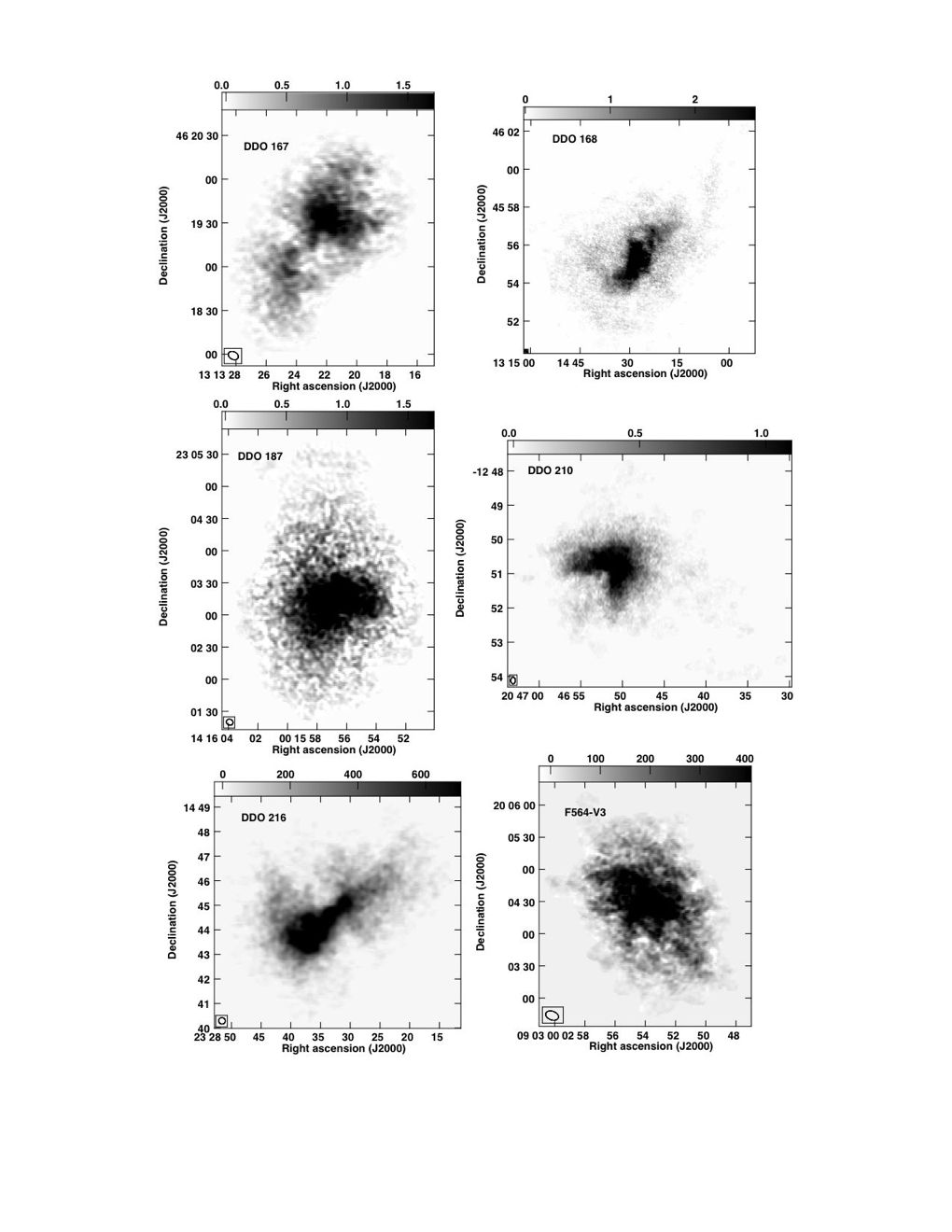}

\clearpage

\plotone{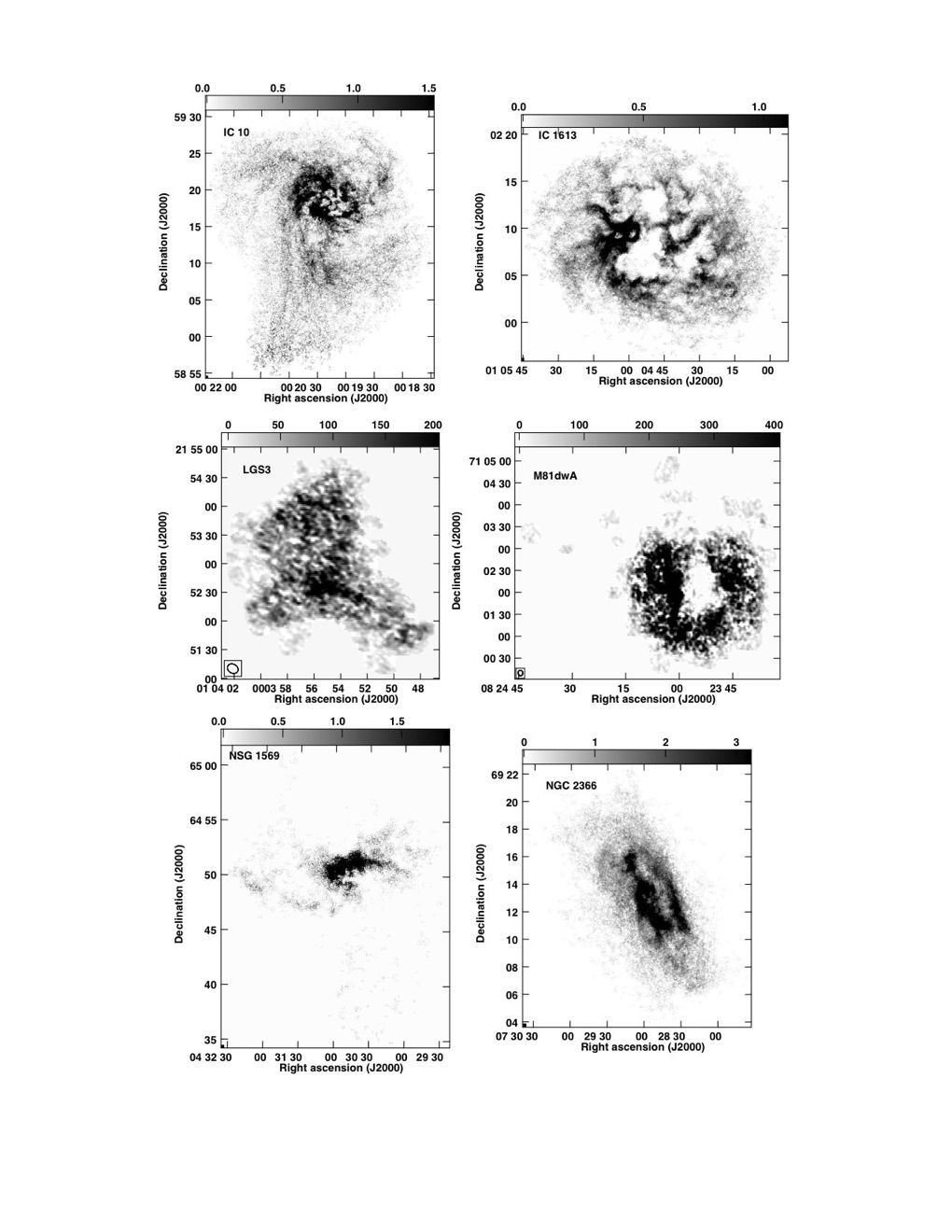}

\clearpage

\plotone{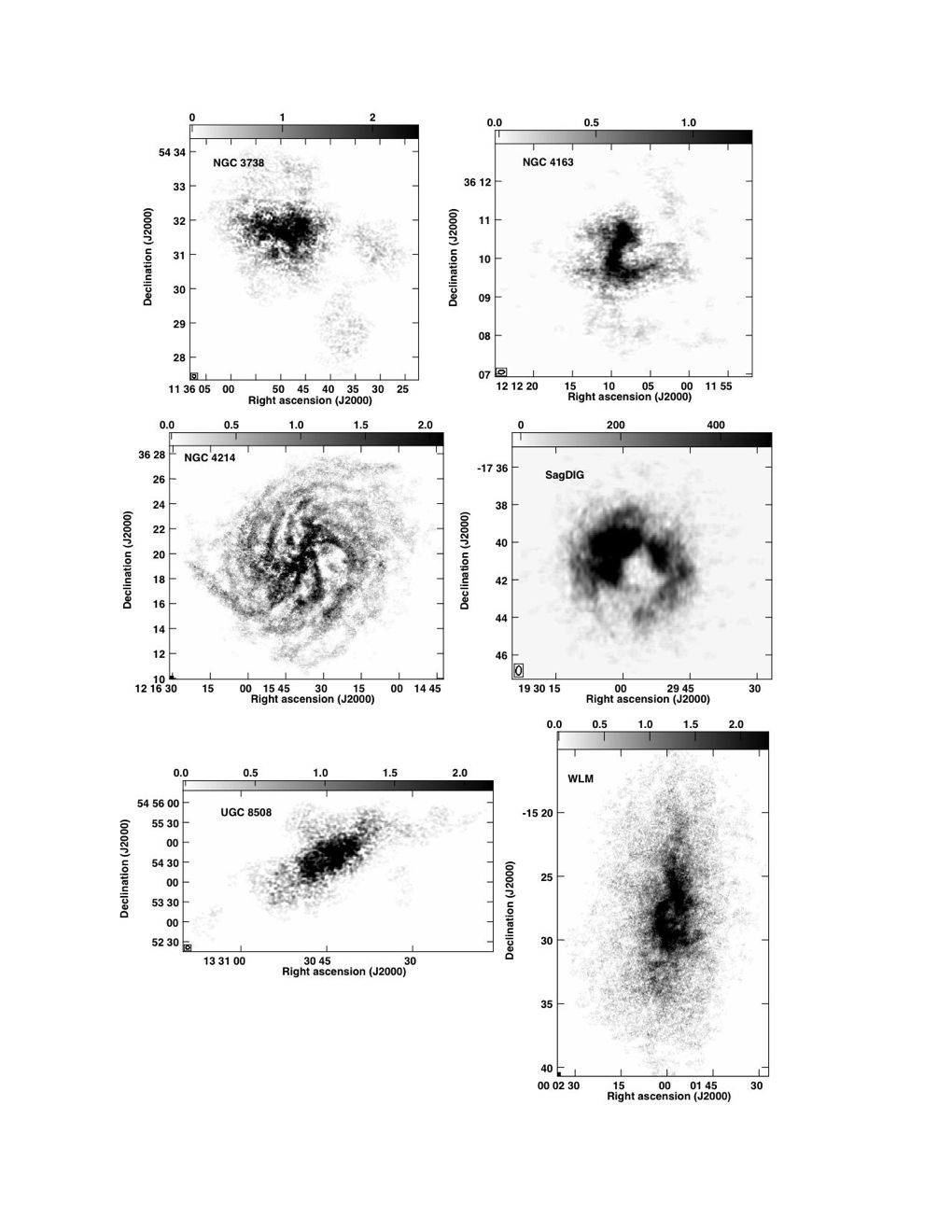}

\clearpage

\plotone{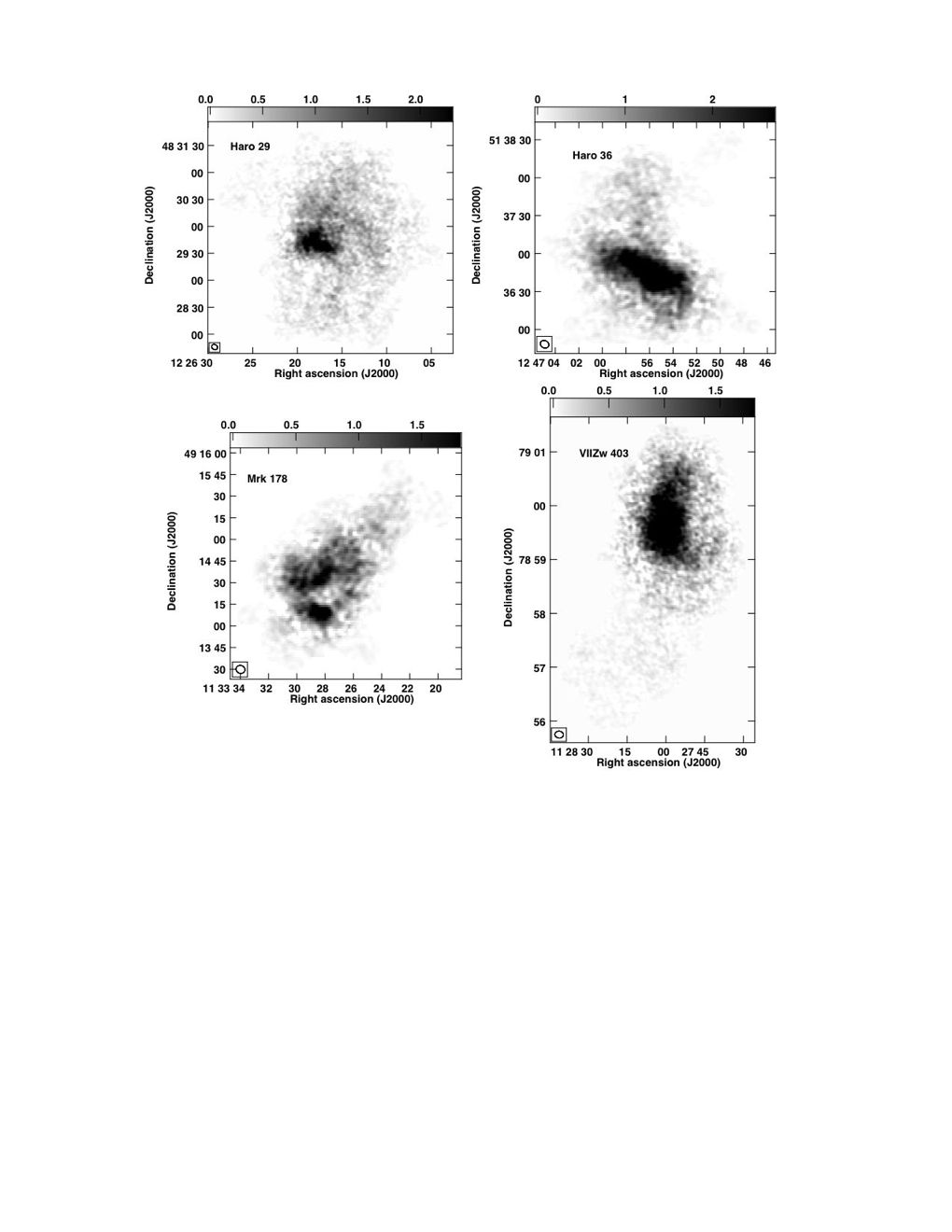}

\clearpage

\clearpage

\plotone{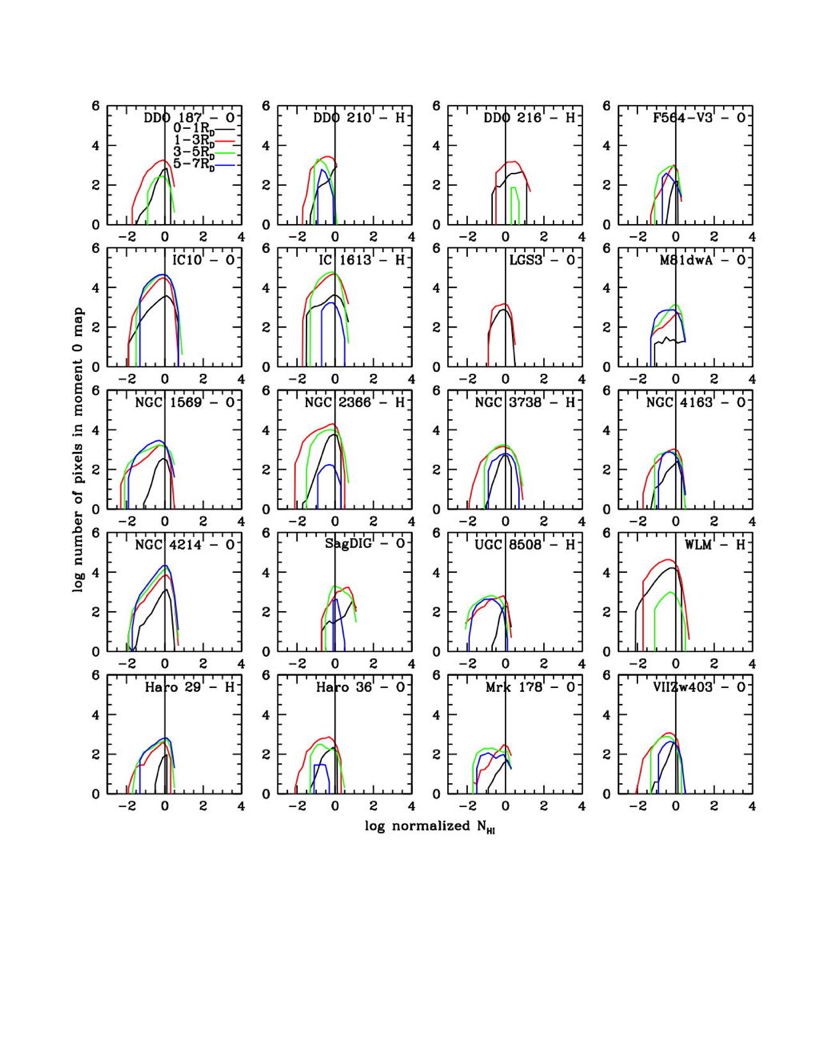}

\clearpage

\clearpage

\plotone{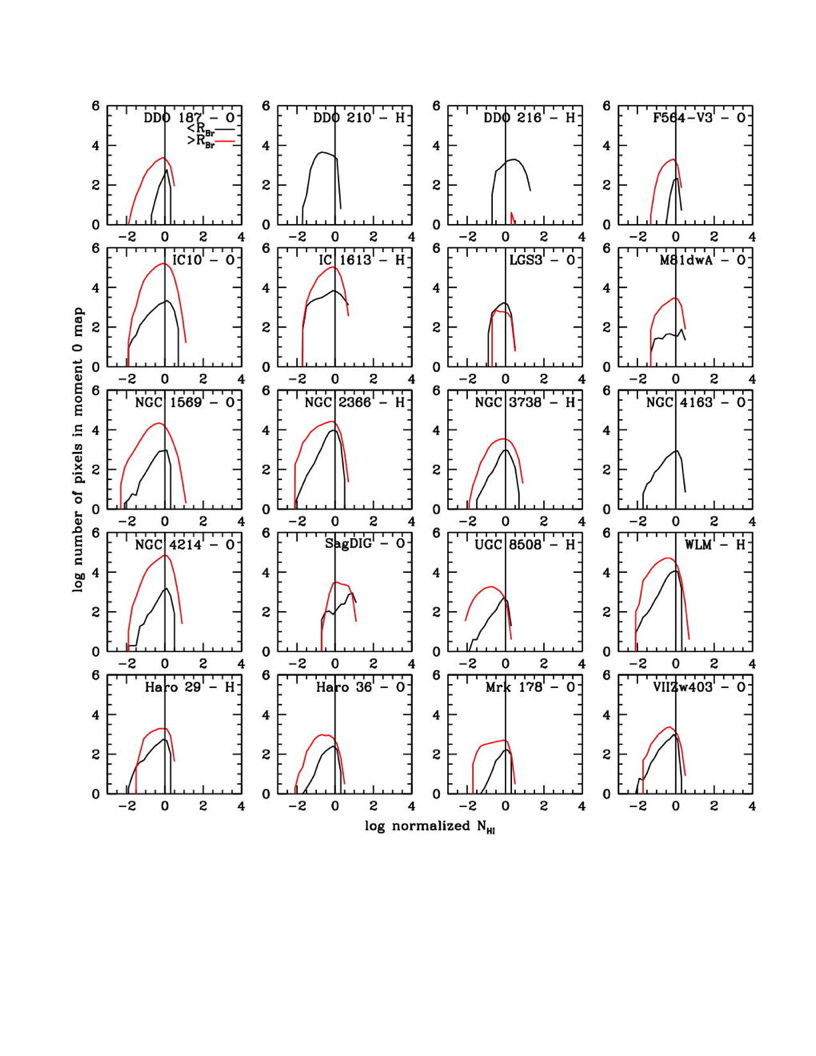}

\clearpage

\clearpage

\clearpage

\clearpage

\clearpage

\clearpage

\clearpage

\clearpage

\clearpage

\clearpage

\clearpage
\epsscale{0.9}
\plotone{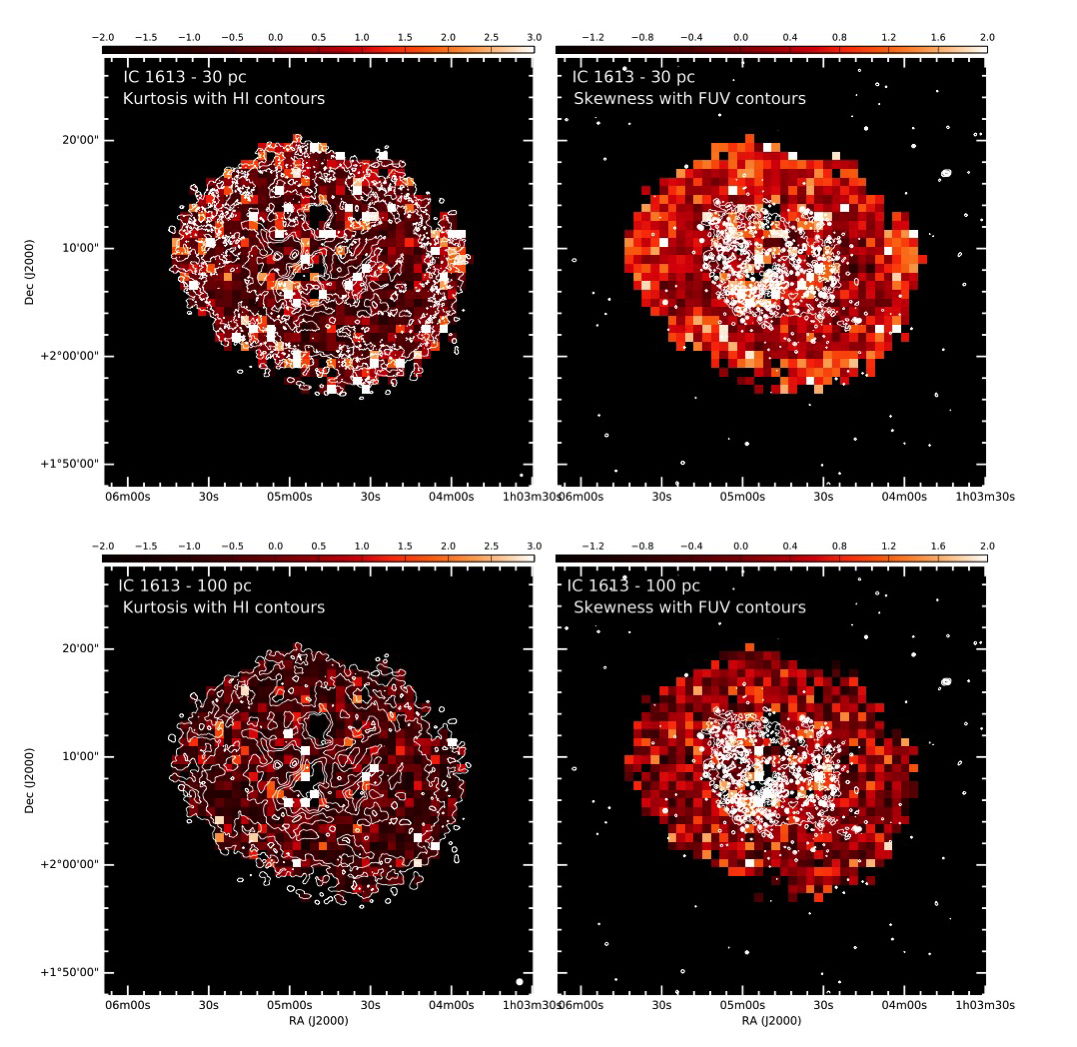}

\clearpage
\epsscale{0.9}
\plotone{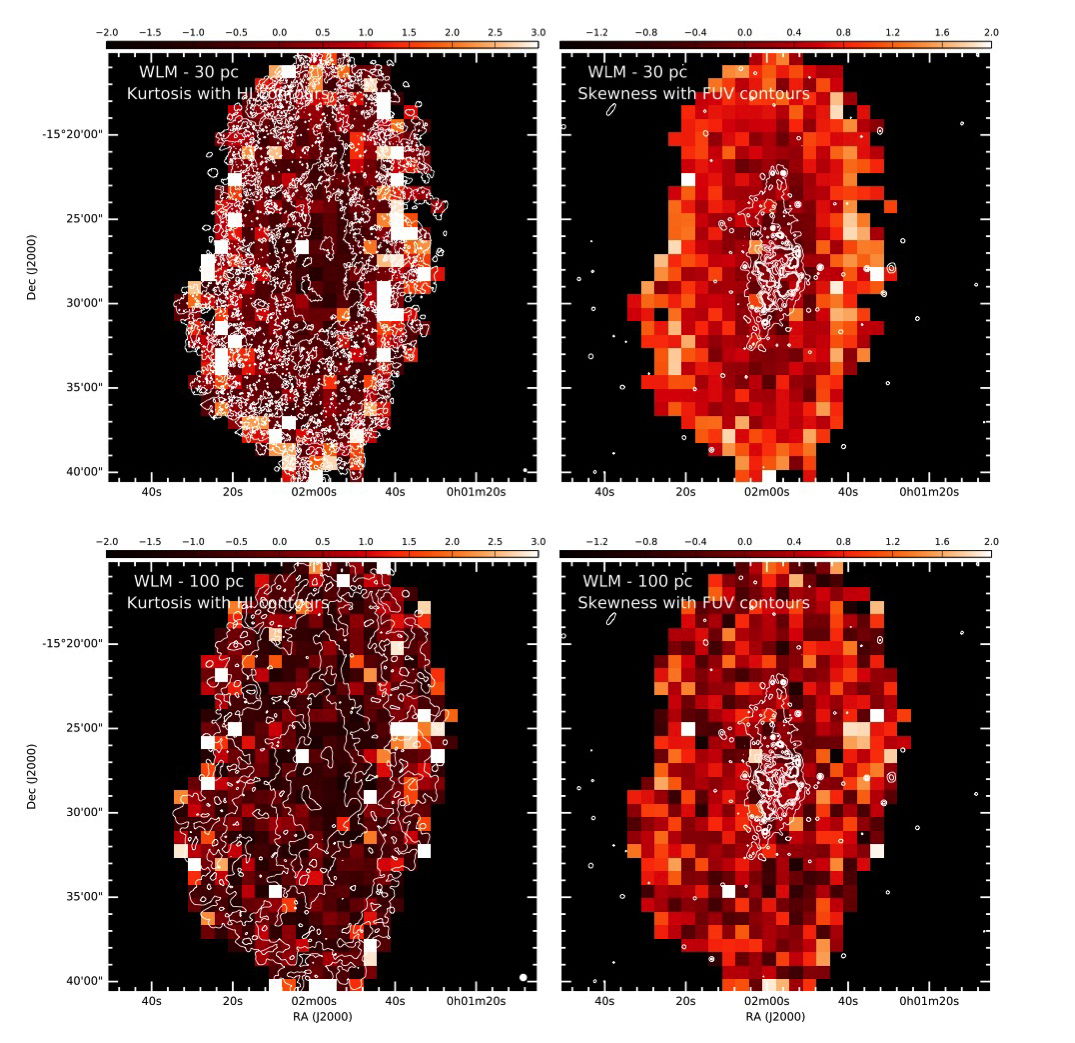}

\clearpage
\epsscale{0.85}
\plotone{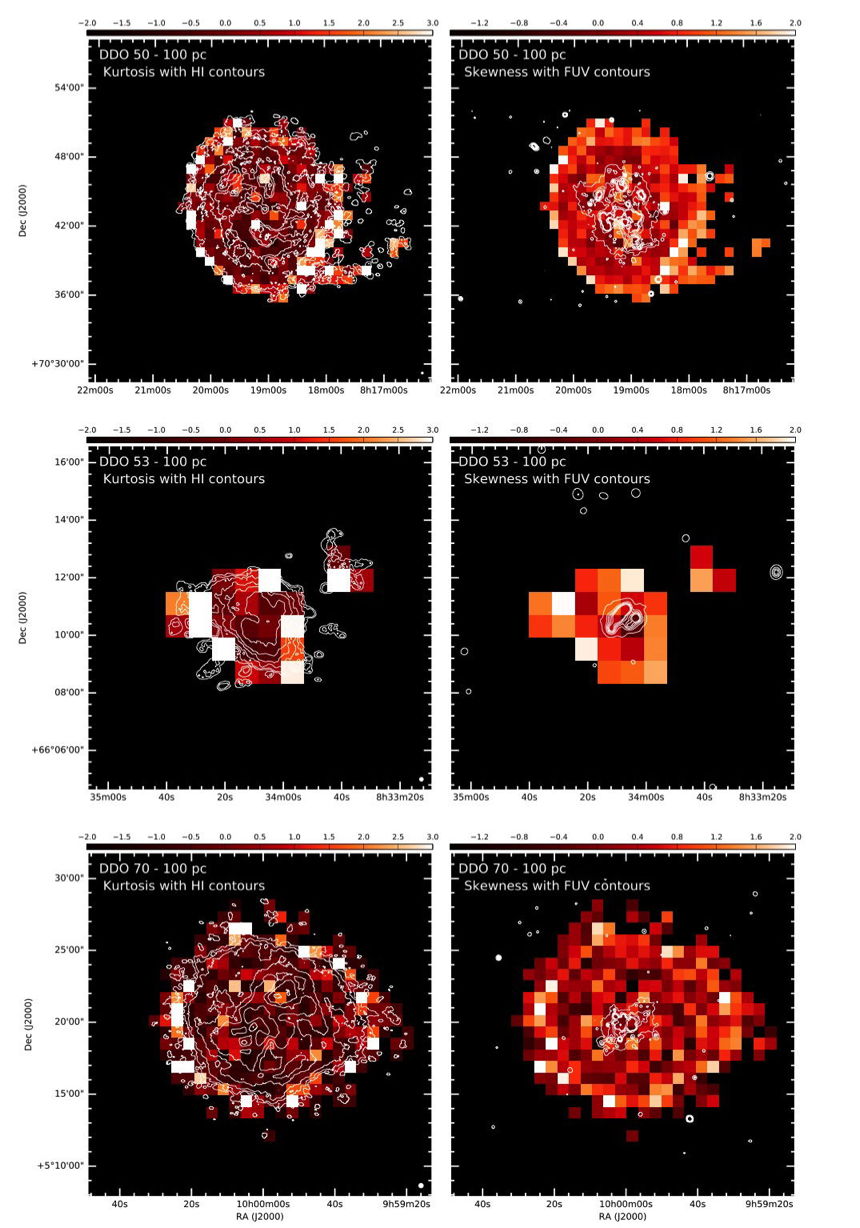}

\clearpage
\epsscale{0.85}
\plotone{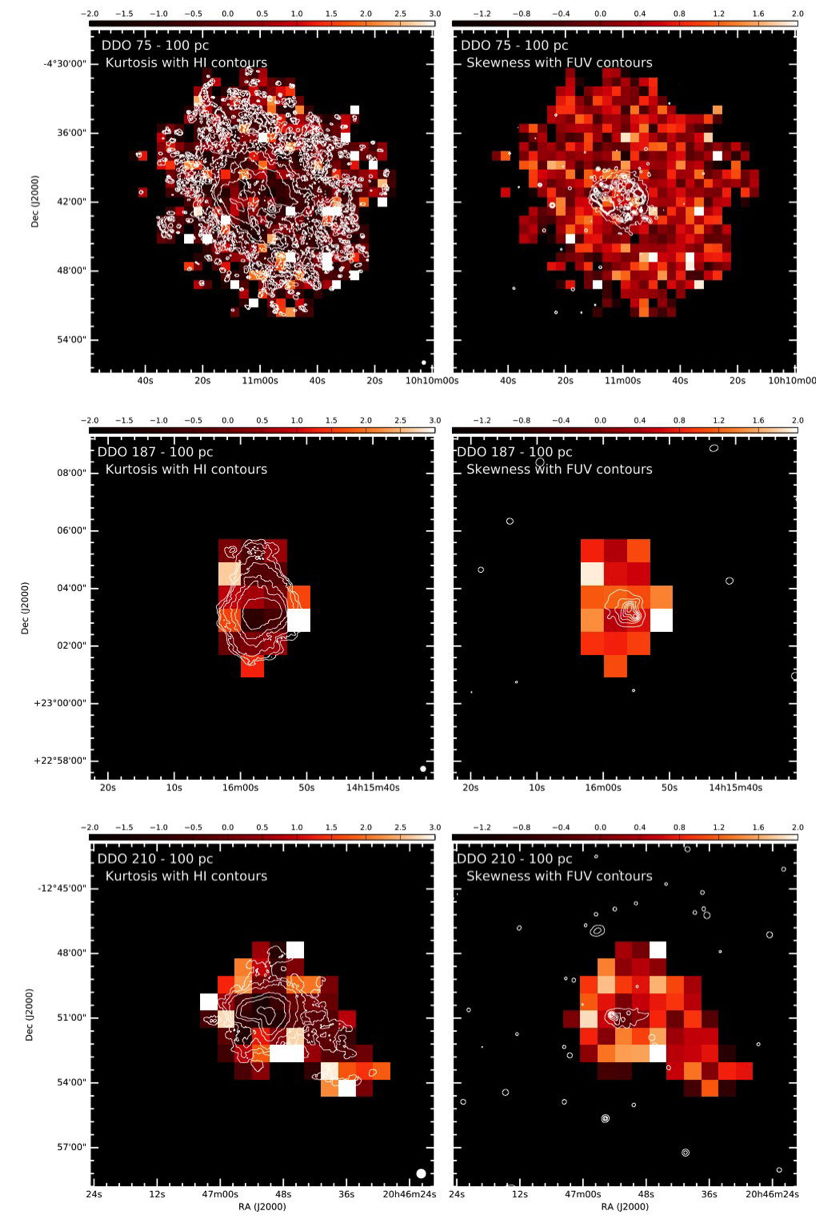}

\clearpage
\epsscale{0.85}
\plotone{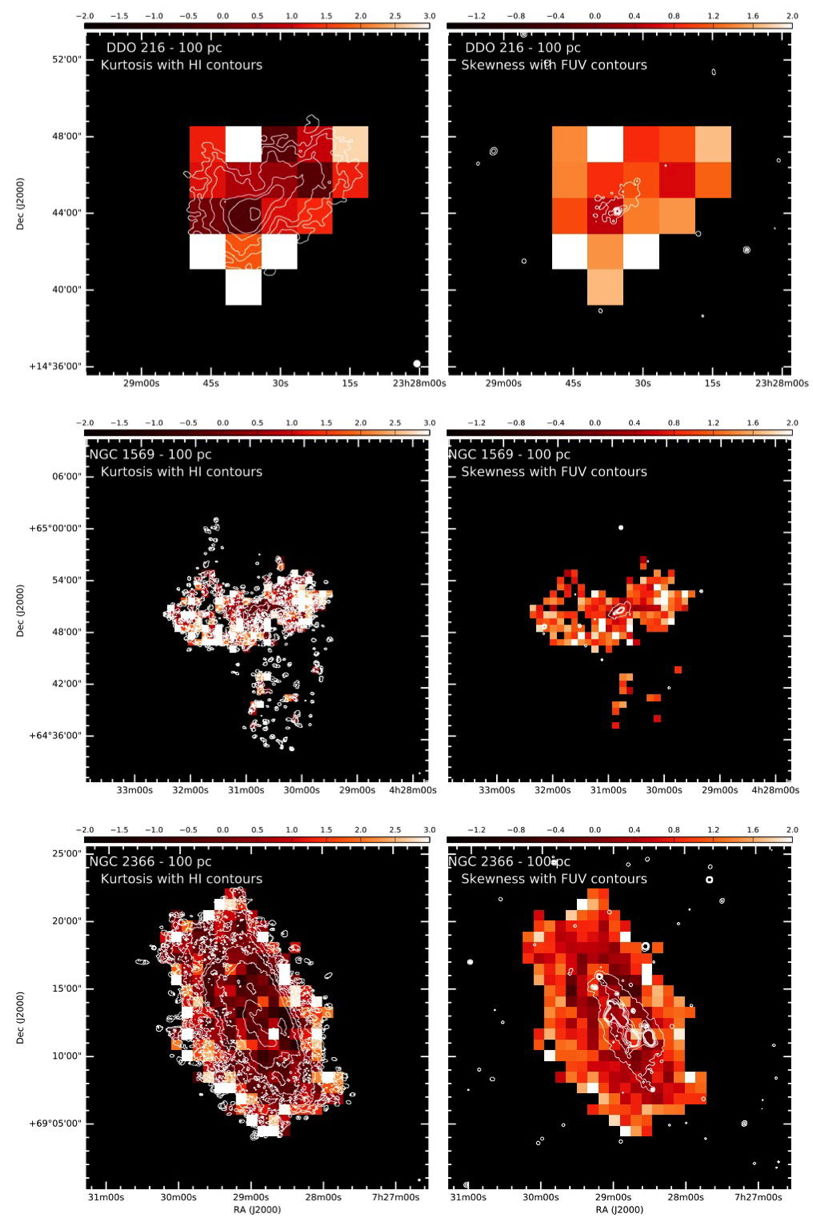}

\clearpage
\epsscale{0.9}
\plotone{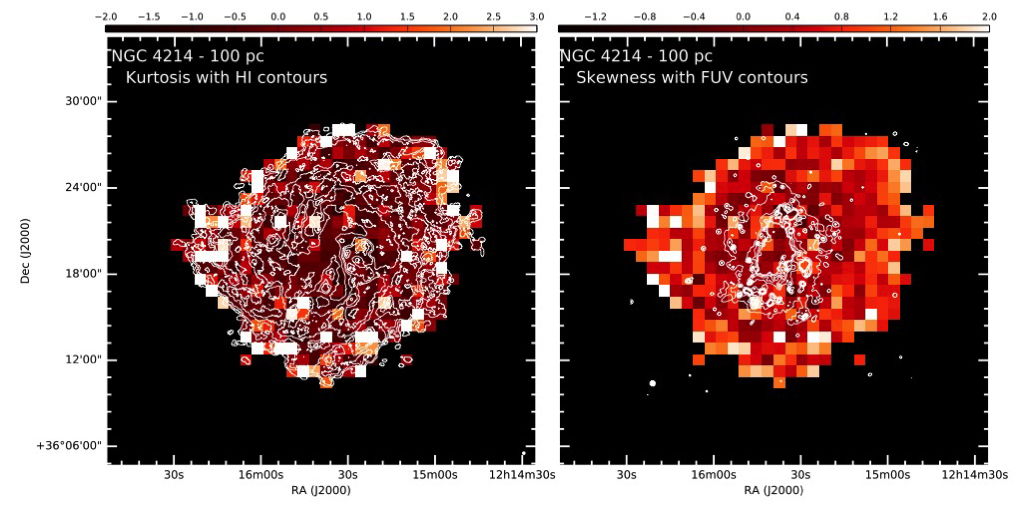}

\clearpage

\clearpage
\epsscale{0.85}
\plotone{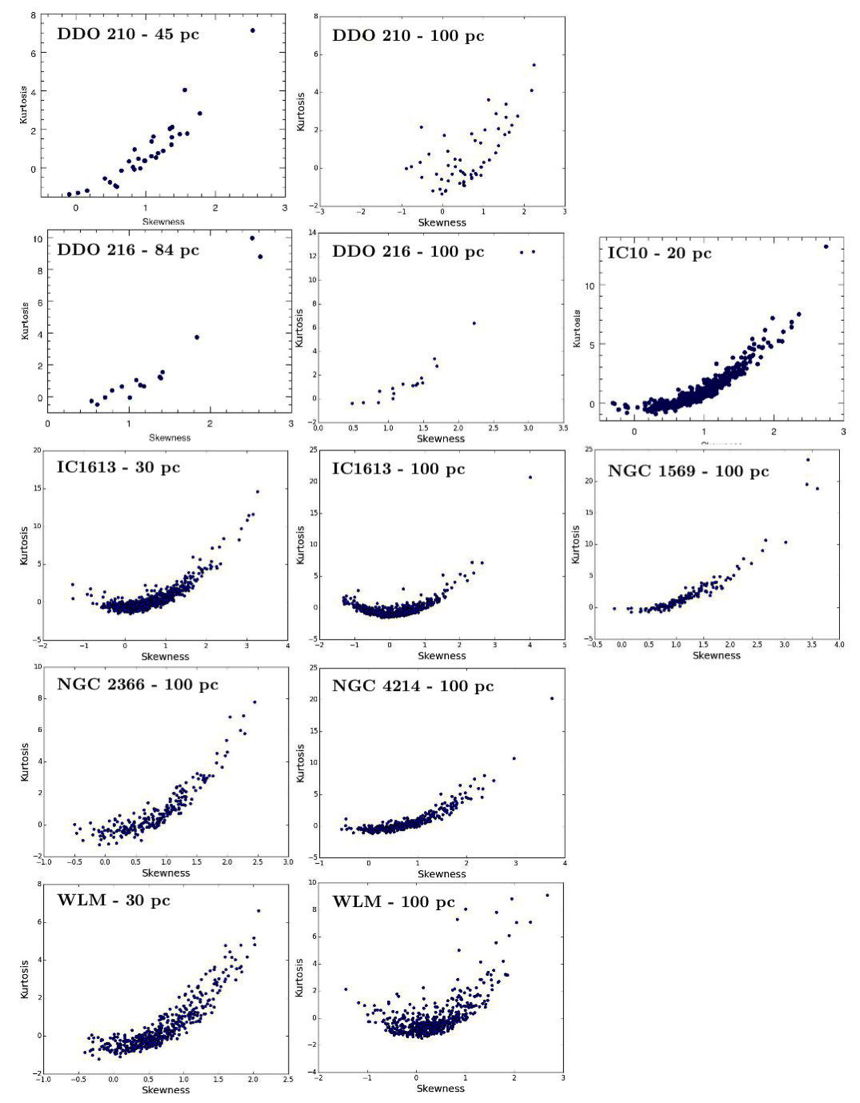}

\clearpage

\begin{deluxetable}{lcccccc}
\tabletypesize{\scriptsize}
\tablewidth{0pt}
\tablecaption{Galaxies and basic data \label{tab-gal}}
\tablehead{
\colhead{} & \colhead{D} & \colhead{$R_D$\tablenotemark{a}} & \colhead{$R_{Br}$\tablenotemark{b}}
& \colhead{} & \colhead{$\log {SFR}_D^{FUV}$\tablenotemark{c}} & \colhead{} \\
\colhead{Galaxy} & \colhead{(Mpc)} & \colhead{(kpc)} & \colhead{(kpc)}
& \colhead{$M_V$} & \colhead{(M\solar\ yr$^{-1}$ kpc$^{-2}$)} & \colhead{log O/H+12\tablenotemark{d}}
}
\startdata
CvnIdwA   &  3.6 & 0.25$\pm$0.12 &  0.56$\pm$0.49  & -12.37$\pm$0.07 & -1.77$\pm$0.42 & $7.3\pm0.06$ \\
DDO 43    &  7.8 & 0.87$\pm$0.10 &  1.46$\pm$0.53  & -15.06$\pm$0.02 & -2.20$\pm$0.10 & $8.3\pm0.09$ \\
DDO 46    &  6.1 & 1.13$\pm$0.05 &  1.27$\pm$0.18  & -14.67$\pm$0.01 & -2.45$\pm$0.04 & $8.1 \pm 0.1$ \\
DDO 47    &  5.2 & 1.34$\pm$0.05 &    \nodata            & -15.46$\pm$0.01 & -2.38$\pm$0.03 & $7.8\pm0.2$ \\
DDO 50    &  3.4 & 1.48$\pm$0.06 &  2.65$\pm$0.27  & -16.61$\pm$0.00 & -1.81$\pm$0.04 & $7.7\pm0.14$ \\
DDO 52    & 10.3 & 1.26$\pm$0.04 &  2.80$\pm$1.35 & -15.45$\pm$0.03 & -2.53$\pm$0.03 & (7.7) \\
DDO 53    &  3.6 & 0.47$\pm$0.01 &  0.62$\pm$0.09  & -13.84$\pm$0.01 & -1.96$\pm$0.02 & $7.6\pm0.11$ \\
DDO 63    &  3.9 & 0.68$\pm$0.01 &  1.31$\pm$0.10  & -14.78$\pm$0.02 & -2.05$\pm$0.02 & $7.6\pm0.11$ \\
DDO 69    &  0.8 & 0.19$\pm$0.01 &  0.27$\pm$0.05  & -11.67$\pm$0.01 & -2.22$\pm$0.05 & $7.4\pm0.10$ \\
DDO 70    &  1.3 & 0.44$\pm$0.01 &  0.13$\pm$0.07  & -14.10$\pm$0.00 & -2.17$\pm$0.02 & $7.5\pm0.06$ \\
DDO 75    &  1.3 & 0.18$\pm$0.01 &  0.71$\pm$0.08  & -13.91$\pm$0.01 & -0.99$\pm$0.05 & $7.5\pm0.06$ \\
DDO 87    &  7.7 & 1.21$\pm$0.02 &  0.99$\pm$0.11  & -14.98$\pm$0.02 & -2.61$\pm$0.02 & $7.8\pm0.04$ \\
DDO 101   &  6.4 & 0.97$\pm$0.06 &  1.16$\pm$0.11 & -15.01$\pm$0.01 & -2.84$\pm$0.05 & $8.7\pm0.03$ \\
DDO 126   &  4.9 & 0.84$\pm$0.13 &  0.60$\pm$0.05 & -14.85$\pm$0.01 & -2.18$\pm$0.13 & (7.8) \\
DDO 133   &  3.5 & 1.22$\pm$0.04 &  2.25$\pm$0.24 & -14.76$\pm$0.01 & -2.60$\pm$0.03 & $8.2\pm0.09$ \\
DDO 154   &  3.7 & 0.48$\pm$0.02 &  0.62$\pm$0.09 & -14.19$\pm$0.01 & -1.77$\pm$0.04 & $7.5\pm0.09$ \\
DDO 155   &  2.2 & 0.15$\pm$0.01 &  0.20$\pm$0.04 & -12.53$\pm$0.02 & \nodata & $7.7\pm0.06$ \\
DDO 165   &  4.6 & 2.24$\pm$0.08 &  1.46$\pm$0.08 & -15.60$\pm$0.01 & \nodata & $7.6\pm0.08$ \\
DDO 167   &  4.2 & 0.22$\pm$0.01 &  0.56$\pm$0.11 & -12.98$\pm$0.04 & -1.59$\pm$0.04 & $7.7\pm0.2$ \\
DDO 168   &  4.3 & 0.83$\pm$0.01 &  0.72$\pm$0.07 & -15.72$\pm$0.01 & -2.06$\pm$0.01 & $8.3\pm0.07$ \\
DDO 187   &  2.2 & 0.37$\pm$0.06 &  0.28$\pm$0.05 & -12.68$\pm$0.01 & -2.60$\pm$0.14 & $7.7\pm0.09$ \\
DDO 210   &  0.9 & 0.16$\pm$0.01 &   \nodata            & -10.88$\pm$0.01 & -2.66$\pm$0.08 & (7.2) \\
DDO 216   &  1.1 & 0.52$\pm$0.01 &  1.77$\pm$0.45 & -13.72$\pm$0.00 & -3.17$\pm$0.02 & $7.9\pm0.15$ \\
F564-V3   &  8.7  & 0.63$\pm$0.09 &  0.73$\pm$0.40  & -13.97$\pm$0.03 & -2.94$\pm$0.13 & (7.6) \\
IC 10         &  0.7 & 0.39$\pm$0.01 &  0.30$\pm$0.04  & -16.34$\pm$0.00 & \nodata & $8.2\pm0.12$ \\
IC 1613     &  0.7 & 0.53$\pm$0.02 &  0.71$\pm$0.12  & -14.60$\pm$0.00 & -1.97$\pm$0.03 & $7.6\pm0.05$ \\
LGS 3       &  0.7  & 0.16$\pm$0.01 &  0.27$\pm$0.08 &    -9.74$\pm$0.03 & -3.75$\pm$0.08 & (7.0) \\
M81dwA    &  3.6 & 0.27$\pm$0.00 &  0.38$\pm$0.03  & -11.73$\pm$0.06 & -2.30$\pm$0.01 & (7.3) \\
NGC 1569  &  3.4 & 0.46$\pm$0.02 &  0.85$\pm$0.24 & -18.24$\pm$0.00 & -0.32$\pm$0.04 & $8.2\pm0.05$ \\
NGC 2366  &  3.4 & 1.91$\pm$0.25 &  2.57$\pm$0.80 & -16.79$\pm$0.00 & -2.04$\pm$0.11 & $7.9\pm0.01$ \\
NGC 3738  &  4.9 & 0.77$\pm$0.01 &  1.16$\pm$0.20 & -17.12$\pm$0.00 & -1.52$\pm$0.02 & $8.4\pm0.01$ \\
NGC 4163  &  2.9 & 0.32$\pm$0.00 &  0.71$\pm$0.48 & -14.45$\pm$0.00 & -1.89$\pm$0.01 & $7.9\pm0.2$ \\
NGC 4214  &  3.0 & 0.75$\pm$0.01 &  0.83$\pm$0.14 & -17.63$\pm$0.00 & -1.11$\pm$0.02 & $8.2\pm0.06$ \\
SagDIG      &  1.1 & 0.32$\pm$0.05 &  0.57$\pm$0.14 & -12.46$\pm$0.01 & -2.40$\pm$0.14 & $7.3\pm0.1$ \\
UGC 8508  & 2.6  & 0.23$\pm$0.01 &  0.41$\pm$0.06 & -13.59$\pm$0.01 & \nodata & $7.9\pm0.2$ \\
WLM          &  1.0  & 1.18$\pm$0.01 &  0.83$\pm$0.16 & -14.39$\pm$0.00 & -2.78$\pm$0.18 & $7.8\pm0.06$ \\
Haro 29   &  5.8 & 0.33$\pm$0.00 &  1.15$\pm$ 0.26  & -14.66$\pm$0.01 & -1.21$\pm$0.01 & $7.9\pm0.07$ \\
Haro 36   &  9.3 & 1.01$\pm$0.00 &  1.16$\pm$ 0.13  & -15.91$\pm$0.00 & -1.88$\pm$0.01 & $8.4\pm0.08$ \\
Mrk 178   &  3.9 & 0.19$\pm$0.00 &  0.38$\pm$ 0.00  & -14.12$\pm$0.01 & -1.17$\pm$0.01 & $7.7\pm0.02$ \\
VIIZw 403 &  4.4 & 0.53$\pm$0.02 &  1.02$\pm$ 0.29 & -14.27$\pm$0.01 & -1.80$\pm$0.03 & $7.7\pm0.01$ \\
\enddata
\tablenotetext{a}{$R_D$ is the disk scale length given by \citet{herrmann13}.}
\tablenotetext{b}{$R_{Br}$ is the radius at which the $V$-band surface brightness profile changes slope, as
given by \citet{herrmann13}.}
\tablenotetext{c}{Integrated SFRs determined from {\it GALEX} FUV fluxes divided by $\pi R_D^2$. The values here differ in some cases
from those given by \citet{lt12} due to a different value for $R_D$. The $R_D$ used here are given in column 3.}
\tablenotetext{d}{See references in \citet{lt12}. Values in parentheses were determined from the empirical
relationship between oxygen abundance and $M_B$
given by \citet{ohmb95} and are uncertain.}
\end{deluxetable}

\clearpage

\begin{deluxetable}{lcccccccccccc}
\tabletypesize{\scriptsize}
\rotate
\tablewidth{0pt}
\tablecaption{Galaxy-wide kurtosis and skewness values \label{tab-results}}
\tablehead{
\colhead{} & \multicolumn{3}{c}{------------Original beam size------------} & \colhead{}
& \multicolumn{2}{c}{------30 pc beam------} & \colhead{}
& \multicolumn{2}{c}{------100 pc beam------} & \colhead{}
& \multicolumn{2}{c}{------200 pc beam------} \\
\colhead{Galaxy} & \colhead{Beam size (pc)} & \colhead{Skewness} & \colhead{Kurtosis} & \colhead{}
& \colhead{Skewness} & \colhead{Kurtosis} & \colhead{}
& \colhead{Skewness} & \colhead{Kurtosis} & \colhead{}
& \colhead{Skewness} & \colhead{Kurtosis}
}
\startdata
CvnIdwA   &  191$\times$183   &  2.28$\pm$0.13 &  4.83$\pm$0.25 &  & \nodata & \nodata &  & \nodata  & \nodata  &  &  2.30$\pm$0.13  &  4.91$\pm$0.27  \\
DDO 43    &  305$\times$225   &  1.63$\pm$0.08 &  2.16$\pm$0.15 &  & \nodata  & \nodata  &  & \nodata  & \nodata  &  & \nodata  & \nodata  \\
DDO 46    &  185$\times$155   &  1.74$\pm$0.06 &  3.36$\pm$0.12 &  & \nodata  & \nodata &  & \nodata  & \nodata  &  & 1.81$\pm$0.07  &  3.55$\pm$0.13  \\
DDO 47    &  263$\times$228   &  1.04$\pm$0.05 &  0.61$\pm$0.11 &  & \nodata  & \nodata  &  & \nodata  & \nodata  &  & \nodata  & \nodata  \\
DDO 50    &  116$\times$101   &  1.70$\pm$0.02 &  3.44$\pm$0.04 &  & \nodata  & \nodata  &  & \nodata & \nodata &  & 1.93$\pm$0.03  &  4.04$\pm$0.07  \\
DDO 52    &  341$\times$259   &  0.97$\pm$0.08 &  0.57$\pm$0.15 &  & \nodata  & \nodata  &  & \nodata  & \nodata  &  & \nodata  & \nodata  \\
DDO 53    &  111$\times$99     &  1.77$\pm$0.07 &  3.03$\pm$0.14 &  & \nodata  & \nodata  &  & \nodata & \nodata &  & 2.42$\pm$0.10 & 5.91$\pm$0.21  \\
DDO 63    & 147$\times$114    &  1.33$\pm$0.05 & 1.25$\pm$0.10  &  & \nodata  & \nodata  &  & \nodata  & \nodata  &  & 1.60$\pm$0.07 & 1.84$\pm$0.14  \\
DDO 69    &   22$\times$21      &  2.50$\pm$0.03 &  7.96$\pm$0.06 &  &  2.83$\pm$ 0.04 & 10.16$\pm$0.08   &  & 3.05$\pm$0.11 & 11.44$\pm$0.23  &  & 1.83$\pm$0.28 & 3.53$\pm$0.57  \\
DDO 70    &   87$\times$83      &  1.77$\pm$0.05 &  3.85$\pm$0.10 &  & \nodata  & \nodata  &  &  1.74$\pm$0.06 & 3.67$\pm$0.12 &  & 1.48$\pm$0.12 & 2.23$\pm$0.24  \\
DDO 75    &   48$\times$41      &  2.87$\pm$0.02 & 11.14$\pm$0.05 &  & \nodata  & \nodata  &  &  3.69$\pm$0.04 & 17.75$\pm$0.08  &  &  2.55$\pm$0.10 & 8.05$\pm$0.20  \\
DDO 87    &  283$\times$232   &  1.32$\pm$0.06 &  2.14$\pm$0.13 &  & \nodata  & \nodata  &  & \nodata  & \nodata  &  & \nodata  & \nodata  \\
DDO 101   &  258$\times$	216  &  0.78$\pm$0.17 & -0.16$\pm$0.34 &  & \nodata  & \nodata  &  & \nodata  & \nodata  &  & \nodata  & \nodata  \\
DDO 126   &  164$\times$	132  &  1.06$\pm$0.07 &  0.54$\pm$0.14 &  & \nodata  & \nodata  &  & \nodata  & \nodata  &  &  1.07$\pm$0.09 & 0.41$\pm$0.18  \\
DDO 133   &  210$\times$183  &  1.04$\pm$0.09 &  0.58$\pm$0.17 &  & \nodata  & \nodata  &  & \nodata  & \nodata  &  & \nodata & \nodata \\
DDO 154   & 142$\times$113   &  2.21$\pm$0.03 &  5.31$\pm$0.06 &  & \nodata  & \nodata  &  & \nodata  & \nodata  &  &  2.54$\pm$0.04 & 6.84$\pm$0.08  \\
DDO 155   & 120$\times$108   &  1.90$\pm$0.13 &  3.08$\pm$0.26 &  & \nodata  & \nodata  &  & \nodata  & \nodata  &  &  1.81$\pm$0.20 &  2.44$\pm$0.41 \\
DDO 165   & 222$\times$168   &  2.21$\pm$0.07 &  5.04$\pm$0.13 &  & \nodata  & \nodata  &  & \nodata  & \nodata  &  & \nodata & \nodata \\
DDO 167   & 148$\times$107   &  1.82$\pm$0.15 &  3.70$\pm$0.30 &  & \nodata  & \nodata  &  & \nodata  & \nodata  &  &  2.08$\pm$0.20 & 4.68$\pm$0.40  \\
DDO 168   & 164$\times$121   &  3.64$\pm$0.04 & 16.34$\pm$0.08 &  & \nodata  & \nodata  &  & \nodata  & \nodata  &  &  4.08$\pm$0.05 & 20.51$\pm$0.10  \\
DDO 187   &   67$\times$59     &  1.79$\pm$0.09 &  2.77$\pm$0.18 &  & \nodata  & \nodata  &  &  2.17$\pm$0.12 & 4.32$\pm$0.25 &  & 2.43$\pm$0.22 &  5.63$\pm$0.44 \\
DDO 210   &   51$\times$38     &  1.98$\pm$0.11 &  3.19$\pm$0.21 &  & \nodata  & \nodata  &  &  2.71$\pm$0.18 & 7.04$\pm$0.37 &  & 1.52$\pm$0.48 & 1.40$\pm$0.97  \\
DDO 216   &   86$\times$82     &  2.28$\pm$0.11 &  5.45$\pm$0.21 &  & \nodata  & \nodata  &  &  2.26$\pm$0.13 & 5.31$\pm$0.25 &  & 2.00$\pm$0.25 & 3.74$\pm$0.50  \\
F564-V3   &  526$\times$342   &  1.48$\pm$0.19 &  1.82$\pm$0.38 &  & \nodata  & \nodata  &  & \nodata  & \nodata  &  & \nodata  & \nodata  \\
IC 10         &   20$\times$19     &  3.18 $\pm$0.01 & 16.15$\pm$0.02 &  & \nodata & \nodata &  & 3.51$\pm$0.05 & 17.97$\pm$0.09 &  & 2.93$\pm$0.09 & 11.33$\pm$0.19  \\
IC 1613     &   26$\times$22     &  3.15$\pm$0.02 &  15.18$\pm$0.03 &  & \nodata & \nodata &  & 2.96$\pm$0.06 & 12.87$\pm$0.12 &  & 2.66$\pm$0.12 & 10.04$\pm$0.24 \\
LGS 3       &   40$\times$32     &  1.14$\pm$0.18 &   1.26$\pm$0.35 &  & \nodata  & \nodata  &  & 0.93$\pm$0.40 &  0.35$\pm$0.80 &  & 0.38$\pm$0.88  & -1.03$\pm$1.77 \\
M81dwA    & 136$\times$110   &  1.15$\pm$0.10 &   0.81$\pm$0.20 &  & \nodata  & \nodata  &  & \nodata & \nodata &  & 1.32$\pm$0.13 & 0.98$\pm$0.26 \\
NGC 1569  &  97$\times$85     &  3.99$\pm$0.03 & 18.84$\pm$0.05 &  & \nodata  & \nodata  &  & \nodata & \nodata &  & 5.70$\pm$0.04 & 37.86$\pm$0.08 \\
NGC 2366  & 114$\times$97    &  2.20$\pm$0.02 &  5.51$\pm$0.05 &   & \nodata  & \nodata  &  & \nodata & \nodata &  & 2.57$\pm$0.04 & 7.35$\pm$0.08  \\
NGC 3738  & 149$\times$131  &  2.82$\pm$0.06 &  9.06$\pm$0.12 &  & \nodata  & \nodata  &  & \nodata  & \nodata &  & 3.20$\pm$0.08 & 11.53$\pm$0.15 \\
NGC 4163  & 136$\times$	83    &  2.37$\pm$0.10 &  5.84$\pm$0.21 &  & \nodata  & \nodata  &  & \nodata  & \nodata &  & 3.28$\pm$0.14 & 11.57$\pm$0.27 \\
NGC 4214  & 110$\times$93    &  2.04$\pm$0.02 &  5.55$\pm$0.04 &  & \nodata  & \nodata  &  & \nodata  & \nodata &  & 2.07$\pm$0.04 & 5.21$\pm$0.07 \\
SagDIG      & 151$\times$90    &  1.80$\pm$0.12 &  3.18$\pm$0.23 &  & \nodata  & \nodata  &  & \nodata  & \nodata &  & 1.66$\pm$0.20 & 2.39$\pm$0.40 \\
UGC 8508  &  74$\times$62     &  2.16$\pm$0.08 &  4.40$\pm$0.16 &  & \nodata  & \nodata  &  & 2.59$\pm$0.10 & 6.46$\pm$0.20 &  & 3.08$\pm$0.16 & 9.46$\pm$0.32 \\
WLM          &   37$\times$25     &  2.29$\pm$0.02 &  5.80$\pm$0.03 &  & \nodata  & \nodata &  & 2.80$\pm$0.05 & 8.62$\pm$0.10 &  & 2.32$\pm$0.11 & 5.39$\pm$0.22 \\
Haro 29     &  192$\times$156   &  3.35$\pm$0.09 & 14.74$\pm$0.17 &  & \nodata  & \nodata  &  & \nodata  & \nodata  &  & 3.47$\pm$0.09 & 15.64$\pm$0.19 \\
Haro 36     &  314$\times$	260   &  2.52$\pm$0.13 &  6.62$\pm$0.26 &  & \nodata  & \nodata  &  & \nodata  & \nodata  &  & \nodata  & \nodata  \\
Mrk 178     &  117$\times$	104   &  1.70$\pm$0.15 &  2.67$\pm$0.30 &  & \nodata  & \nodata  &  & \nodata & \nodata &  & 1.97$\pm$0.21 & 3.24$\pm$0.42  \\
VIIZw 403  &  201$\times$164   &  2.45$\pm$0.12 &  6.26$\pm$0.24 &  & \nodata  & \nodata  &  & \nodata  & \nodata  &  & \nodata & \nodata \\
\enddata
\end{deluxetable}

\clearpage

\begin{deluxetable}{lc}
\tabletypesize{\scriptsize}
\tablewidth{0pt}
\tablecaption{Kernel Analysis \label{tab-kernels}}
\tablehead{
\colhead{Galaxy} & \colhead{32-pixel kernel length (pc)}
}
\startdata
DDO 50    &    790 \\
DDO 53    &    840 \\
DDO 69    &    190 \\
DDO 70    &    300 \\
DDO 75    &    300 \\
DDO 155   &    510 \\
DDO 187   &     510 \\
DDO 210   &     210 \\
DDO 216   &     600 \\
IC 10         &      163 \\
IC 1613     &      163 \\
NGC 1569  &    790 \\
NGC 2366  &    790 \\
NGC 4214  &   700 \\
WLM          &     230 \\
\enddata
\end{deluxetable}

\end{document}